\documentclass[prd,twocolumn,showpacs,nofootinbib,preprintnumbers,superscriptaddress]{revtex4}
\usepackage{amsmath}\usepackage{amsfonts}\usepackage{graphicx}\usepackage{hyperref}\usepackage{mathbbol}
\def\be{\begin{equation}}   \def\ee{\end{equation}}   \def\bea{\begin{eqnarray}}    \def\eea{\end{eqnarray}}  \def\no{\nonumber}
    \def\d{{\rm d}}         \def\k{\kappa}    \def\a{\alpha}   \def\b{\beta}
\def\r{\right}            \def\l{\left}
\def\f{\frac}    \def\l{\left}   \def\r{\right}

\begin{document}
\title{Holographic dark energy with non-minimal derivative coupling to gravity effects}
\date{\today}

\author{Chonticha Kritpetch}\email{chontichakr57@email.nu.ac.th}
 \affiliation{The Institute for Fundamental Study ``The Tah Poe Academia Institute", Naresuan University, Phitsanulok 65000, Thailand}
 \author{Candrasyah Muhammad}\email{candrasyahm59@email.nu.ac.th}
 \affiliation{The Institute for Fundamental Study ``The Tah Poe Academia Institute", Naresuan University, Phitsanulok 65000, Thailand}
\author{Burin Gumjudpai}\email{Corresponding: buring@nu.ac.th}
 \affiliation{The Institute for Fundamental Study ``The Tah Poe Academia Institute", Naresuan University, Phitsanulok 65000, Thailand}
\affiliation{Thailand Center of Excellence in Physics, \\ Ministry of Higher Education, Science, Research and Innovation, Bangkok 10400, Thailand}

\begin{abstract}
Non-minimal derivative coupling (NMDC) to gravity in flat FLRW universe is investigated in the scenario of holographic dark energy. Kinetic term is coupled to the Einstein tensor under potential $V = (1/2)m^2 \phi^2$. The free kinetic term is allowed to be canonical and phantom.  Gravitational constant is modified with the NMDC coupling.
Holographic cutoff at Hubble horizon gives modification to dark energy density. We evaluate dark energy equation of state and the variation of gravitational constant of the theory such that the theory can be constrained. It is found that positive NMDC coupling is favored rather than the negative one. The model with purely NMDC theory and the potential is favored with positive sub-Planckian NMDC coupling and decaying scalar field. The canonical scalar field with positive NMDC coupling under the scalar potential is also viable under some conditions that result in oscillating scalar field. The phantom field case is not favored in this model since the coupling and scalar mass are required to be super-Planckian while it is tightly constrained by gravitational constant variation observations.
\end{abstract}
\pacs{98.80.Cq}

\date{\today}

\vskip 1pc

\maketitle \vskip 1pc
\section{Introduction}  \label{sec_intro}
In the past decades, many theoretical approaches to explain cosmological problems, namely present acceleration with $w\simeq -1$ \cite{Amanullah2010, Astier:2005qq, Goldhaber:2001a, Perlmutter:1997zf, Perlmutter:1999a, Riess:1998cb, Scranton:2003, Tegmark:2004}, dark matter and inflationary graceful exits with acceptable CMB data \cite{WMAP9, Aghanim:2019ame}, have been proposed (as reviewed in \cite{VF,CL,CF,odin,en,Ishak:2018his} and many references therein). Two approaches are to modify matter Lagrangian and to modify gravitational sector. In modifying matter sector, dark energy density is added into consideration in form of cosmological constant or various models of scalar field. In modifying gravitational sector, the $f(R)$ theories are to modified Einstein-Hilbert Lagrangian with function of Ricci scalar, Ricci tensor or Riemann tensor \cite{Carroll2004}.
As we allow coupling between barotropic matter to scalar field, chameleon mechanism could safe the theory from solar system fifth force constraints \cite{charm}.
It is possible that both matter and gravity can be modified together at the same time in scalar-tensor theory as coupling between scalar and gravitational sector is allowed \cite{BD1961, Maeda, VF}.

Coupling function in form of $f(\phi, \phi_{,\mu}, \phi_{,\mu\nu}, \ldots)$ was introduced as an modification of the scalar-tensor theory. The term is motivated
in the context of scalar quantum electrodynamics or in gravitational theory of which Newton's constant is a function of the density \cite{Amendola1993} which later was considered as a function of kinetic energy density of the scalar field. In maintaining the action, acceleration solution must be possible.
Other coupling terms apart from
 $R \phi_{,\mu}\phi^{,\mu}$
and  $R^{\mu\nu} \phi_{,\mu} \phi_{,\nu}$ was shown to be unnecessary \cite{Capozziello:1999xt}.
The two terms are of the lower energy limits of higher dimensional
theories or Weyl anomaly of $\mathcal{N} = 4$ conformal supergravity
\cite{Liu:1998bu, Nojiri:1998dh}, hence giving good motivation to the theory.
The non-minimal derivative coupling to gravity terms have been considered as  $\kappa_1 R \phi_{,\mu}\phi^{,\mu}$  and
 $\kappa_2 R^{\mu\nu} \phi_{,\mu} \phi_{,\nu}$. It  usually results in de-Sitter expansion \cite{Capozziello:1999uwa}. These coupling terms were further modified with motivation that the static fixed point suggests  $\kappa \equiv   \kappa_2  =  -2 \kappa_1$
 \cite{Granda:2010hb, Granda:2010ex, Granda:2011zk}. This gives a hint to a theory of NMDC term coupling  to the Einstein tensor, $G^{\mu\nu}$  as both the metric and Einstein tensor are divergence-free making the theory naturally settle    \cite{Sushkov:2009, Saridakis:2010mf, Gao:2010vr, Germani:2010gm, Sushkov:2012, Skugoreva:2013ooa, Koutsoumbas:2013boa, Darabi:2013caa, Germani:2009, Germani:2010hd, Dalianis:2016wpu, Sadjadi:2012zp, Tsujikawa:2012mk, Ema:2015oaa, Jinno:2013fka, Ema:2016hlw, Yang:2015pga, Sadjadi:2010bz, Gumjudpai:2015vio}. Attempts in generalizing scalar-tensor theory with at most second-order derivative with respect to its dynamical variables, i.e. the metric tensor and scalar field, brought about other upgraded versions of the theory, for examples, galileons \cite{Nicolis:2008in, Deffayet:2009wt, Deffayet:2009mn} and the Fab-Four \cite{Charmousis:2011bf}.    Recent literatures found that the NMDC term is a sub-class of the Horndeski action proposed a few decades ago \cite{Horn, Deffayet:2011, Kobayashi} and of the beyond Horndeski action such as GLPV theories \cite{GLPV}.

Considering NMDC theory, the sign of the coupling and its constancy (or variation \cite{Granda:2010hb}) are significantly important since
the NMDC effect could enhance or dilute the effects of the free kinetic term. This could affect power spectral index, tensor-to-scalar ratio, evolution of the equation of state, future Big Rip singularity and its phantom crossing \cite{Caldwell:2003vq, Saridakis:2010mf, Skugoreva:2013ooa, Quiros:2017gsu}. The theoretical predictions are confronted with observational data leading to tight constraints for particular types of scalar potential \cite{Tsujikawa:2012mk, Yang:2015pga}. It also suggests us if the NMDC model should be suitable either as an inflationary model or as a dark energy model.
The NMDC theory can also be considered in Palatini approach when considering the connection field as an additional dynamical variable,\cite{Gumjudpai:2016ioy, Saichaemchan:2017psl, Muhammad:2018dwi}.

In this work, we consider NMDC theory with constant coupling $\kappa$. Typically the theory is viable under some restricted conditions.
In such theory, quasi-de Sitter expansion can be achieved with graceful exits if without scalar potential $V(\phi)$ and if having fast-roll initial field speed. This is because the positive coupling NMDC theory with a potential but without barotropic fluid allows unbound value of $\dot{\phi}$ but finite value of Hubble parameter. The equation of state of the NMDC theory approaches $-1$ at late time  \cite{Sushkov:2009, Bruneton:2012zk}. Inclusion of constant potential (or cosmological constant) can also result in quasi-de Sitter expansion \cite{Sushkov:2012}.
When including of power-law potential (without barotropic matter), $V(\phi) = V_0 \phi^n$, it is found that acceleration is possible for $n \leq 2$ for a range of scalar mass, $m_{\phi}$ and $\k$, i.e. $\k < G$ and $V_0 < G$ (sub-Planckian), where $G$ is the Newton gravitational constant. The acceleration comes to an end with scale factor oscillation whereas the case $n > 2$ results in Big Rip singularity. The case $n<2$ results in the same inflationary regime as the constant potential case \cite{Skugoreva:2013ooa}. The analysis is extended to include Higgs-like potential and exponential potential in \cite{Matsumoto:2017gnx, Granda:2017dlx}.  The theory produces larger tensor-to-scalar ratio and no graceful exits for $n \leq 2$, hence it should be considered as dark energy rather than inflation in the early universe.

Here we apply the holographic dark energy framework to the NMDC theory. We will examine the NMDC effects on the holographic dark energy density and predict the evolution of equation of state coefficient, $w_{\rm DE}$ and its effects on gravitational constant variation.
There are some works previously considering properties of holographic superconductors with charged scalar field non-minimal derivative coupling (with derivative operator $D_{\mu} = \partial_{\mu}  -  i A_{\mu} $) to Einstein tensor in presence of electromagnetic gauge field in the AdS background \cite{Chen:2010hi, Kuang:2016edj}. Our consideration is different from this one that our derivative operator is covariant derivative and the scalar field is not charged.  As final aim of theoretical physics, quantum gravity is believed to be the theory of unification of all forces and it is yet to complete. However, even without the complete theory of quantum gravity, one can consider the nature of dark energy under some principles in
quantum gravity - the holographic principle \cite{Hooft, Susskind}. Idea of holographic energy is considered as an infrared cutoff which puts limit to the dark energy density \cite{Cohen, Horava, Thomas, Hsu, Li, Pavon}. The UV energy scale, IR length scale, and  entropy are hypothesized to relate as $\rho_{\Lambda} \propto S L^{-4}$ for a blackhole. Blackhole's entropy scales with area as of the Bekenstein-Hawking entropy, $S \sim A/4G \sim L^2$  \cite{Be1, Be2, Hawking, Hawking2}. The  holographic dark energy is an application of this idea, not to blackholes' horizon, but to cosmological length scale, e.g. to the apparent horizon or to the Hubble horizon.
In this framework, the total dark energy in a considered volume can not exceed blackhole mass of the same volume size.  The energy density of holographic dark energy is given hence by
\begin{equation}
\rho_{\Lambda} = \frac{3c^2}{8\pi G L^2},    \label{1}
\end{equation}
where $L$ is infrared cutoff scale and here it is considered as the size of $H^{-1}$. The factor $c$ is a constant.

      In this work, we study holographic dark energy which is scalar field with non-minimal derivative coupling (NMDC) to gravity in spatially flat FLRW universe\footnote{Consideration of non-minimal coupling (NMC) theory as holographic dark energy was studied previously as in \cite{Setare:2008pc}.}. The $w_{\rm DE} = 0$ problem of the holographic dark energy model in flat FLRW universe \cite{Hsu} is solved with the NMDC modification. The other type of IR cutoffs are reviewed in \cite{Wang:2016och}. One type is  future event horizon,
 $     R_{h} = a \int^{\infty}_{t}  a^{-1}  {\d t'} = a \int^{\infty}_{a}   (1/{Ha'^{2}}) {\d a'}  $  which resolves  the cosmic coincidence
 problem but  violating  causality  \cite{Li}. This is  since the future even horizon is used to
determine present day acceleration. It is as well  found that this model failed to predict age of the universe  \cite{Wei:2007ig}.
Another model is  agegraphic dark energy of which the IR cutoff is chosen to be conformal time \cite{Cai:2007us, Wei:2007ty,Wei:2007xu}. In this model, energy density of dark energy is taken to be  $
\rho_{\rm DE} = {3n^{2} (8 \pi G)}/{\eta^{2}}, $
where $\eta$ is  conformal time
$ \eta = \int {\d t}/{a} = \int  ({a^{2}H})^{-1} {\d a}$.  This model can solve cosmic coincidence problem, but strongly disfavored by data.
   The   Ricci scalar can also be considered as the IR cutoff. The Ricci dark energy density is proportional to Ricci scalar  \cite{Gao:2007ep},
$
\rho_{\rm DE} = - {\alpha R}/{(16\pi)} = \l[{3\alpha}/({8\pi})\r] \left(\dot{H} + 2H^{2}+ {k}/{a^{2}} \right). $
Observational constraints for Ricci dark energy are shown  in \cite{Zhang:2009un}. The other  IR cut-off is proposed by Granda and Oliveros  \cite{Granda:2008dk}\cite{Granda:2008tm} with dark energy density, $
\rho_{\Lambda} \approx \alpha H^{2} + \beta\dot{H}$   where $\alpha$ and $\beta$ are model parameters, therefore it includes the Ricci dark energy model. In some case, higher-order derivative of the Hubble parameter could be considered in the cutoff effect, these are such as in
\cite{Nojiri:2005pu, Chen:2009zzv, Chattopadhyay:2020mqj, Chattopadhyay:2020xx, Chattopadhyay:2020x2}.
Currently this model is still widely studied in the context of some modified gravity theories.  Different forms of the cutoff  would change the character of the Friedmann equation dramatically.

 It  was  recently shown that for the case of $\k > 0$, the NMDC theory has classical (Laplacian) instability, i.e. square of sound speed, $c_{\rm s}^2$ is negative\footnote{The sign of coupling constant in \cite{Quiros:2017gsu} is defined oppositely from our convention here. Moreover, it is noted that the Klein-Gordon equation ($\ddot{\phi}$ equation) in \cite{Quiros:2017gsu} was reported differently from ours. This leads to different expression of $p_{\phi}$ which results in different predictions of $c_{\rm s}^2$. Hence the mentioned shortcomings of the $\k > 0$ case is not yet to conclude.} \cite{Quiros:2017gsu}. However we notice that $c_{\rm s}^2$ of the theory is a function of $\dot{\phi}, \ddot{\phi}, H$ and $\dot{H}$ hence different dynamics could change the range of $c_{\rm s}^2$ significantly.  The holographic modification of the dark energy density modifies $\rho_{\Lambda}$ term which, as a result, modifies NMDC kinetic term in the equations of motion. Hence it should significantly change Laplacian instability.

Motivations of both NMDC and holographic dark energy can be well fit in quantum gravities such as string theory and both do not contradict to each other. There are no reason that such scalar-tensor theories can not be incorporated with holographic ideas. The two dark energy ideas can be considered as one picture in describing the late acceleration.
 In section II, NMDC gravity and its cosmological aspects of the theory are introduced and discussed. In section III, we discuss the idea of holographic dark energy and draw the NMDC effect into the scenario of holographic dark energy. We investigate the situation when there is only NMDC term damping the universe without the free kinetic term  in section IV. In section V, both kinetic terms are considered together. Equations of state are analyzed in each case of canonical and phantom scalar fields. Constraints from variation of gravitational constant are considered in section VI. At last, the conclusion and critics are given in section VII.

\section{Non-minimal derivative coupling (NMDC) gravity}
We consider gravitational action in form of
\be
S = \int\mathrm{d}^4x\sqrt{-g}\left\{ \frac{R}{16 \pi G}   - \f{\left[ \varepsilon g_{\mu\nu}+  \k G_{\mu\nu}     \right]  }{2} (\nabla^{\mu} \phi)(\nabla^{\nu} \phi) -  V  \right\},      \label{action}
\ee
together with matter action $S_{\rm m}$ where  $G_{\mu\nu}$ is the Einstein tensor,  $\varepsilon = \pm 1$ for canonical and phantom case, or $\varepsilon = 0$ when considering purely NMDC kinetic term. $V=V(\phi)$ is scalar potential. Factor $\k$ is the NMDC coupling constant \cite{Granda:2010ex, Sushkov:2009, Saridakis:2010mf}.
The Lagrangian is a sub-class of Horndeski theory \cite{Horn, Kobayashi, Deffayet:2011, Charmousis:2011bf}
 with $
G_2  =  -({\varepsilon}/{2}) g_{\mu\nu}(\nabla^{\mu} \phi)(\nabla^{\nu} \phi),\:  G_3 = 0, \: G_4 = (16 \pi G)^{-1},\:     G_5 = c_5 \phi = \phi {\k}/{2},  \: {\text{with}}   \: c_5 \equiv {\k}/{2}$.  Here $H$ is the Hubble parameter and $\rho_{\rm m}, p_{\rm m}$ are the energy density and pressure of
matter. For spatially flat FLRW universe as
\begin{equation}
\mathrm{d}s^2 = -\mathrm{d}t^2+a(t)^2(\mathrm{d}r^2+r^2 \mathrm{d}\Omega^2)\,.
\end{equation}
The Friedmann equation is viewed in two perspectives.  Firstly, keeping Newton's gravitational constant $G$ standard and having scalar kinetic term modified \cite{Sushkov:2009},
\be
  H^2  =   \f{8\pi}{3}G \left[\frac{1}{2}\dot{\phi}^2(\varepsilon-9\k H^2) + V  +  \rho_{\rm tot}   \right]\,,   \label{fggg}
\ee
where $\rho_{\rm tot}  = \rho_{\rm m}  + \rho_{\Lambda}$.   Secondly, keeping scalar kinetic term in standard form and let the gravitational constant $G$ modified, the equation 
(\ref{fggg}) can be rearranged as 
\be
                    H^2    =   \f{8\pi}{3}G_{\rm eff} \left(\frac{\varepsilon }{2}\dot{\phi}^2 + V  +  \rho_{\rm tot} \right)\,,     \label{5}
\ee
where effective gravitational constant can be read off from the Friedmann equation,
\be
G_{\rm eff}(\dot{\phi})   \equiv  \f{G}{1 + 12 \pi G \k \dot{\phi}^2}\;.   \label{Geff}
\ee
 The other field equation ($ii$ components) is expressed  in the view of modified scalar kinetic term as
\bea
& & 2\dot{H} + 3 H^2    = \no
 \\ && 8 \pi G \left\{-\frac{ \dot{\phi}^2}{2} \left[\varepsilon+\k\left(2\dot{H}+3H^2+4H\frac{\ddot{\phi}}{\dot{\phi}} \right)\right] +  V
  -  p_{\rm tot} \right\},  \no \\  \label{lolo}
\eea
 or in the view of modifying of $G$ and of  modifying energy density together as
\be
 2\dot{H} + 3 H^2    =    \f{ 8 \pi G}{(1 + 4 \pi G \k \dot{\phi}^2)} \left[ - \f{\dot{\phi}^2}{2}\varepsilon   +  V -  p_{\rm tot} - 2 \k  H {\dot{\phi}}{\ddot{\phi}} \right]\,, \label{lolo2}
\ee
where $p_{\rm tot} = p_{\rm m} + p_{\Lambda} $.   These  are combined with the Friedmann equation to give
\be
\dot{H} = - 4\pi G\l[ \dot{\phi}^2 \l( \varepsilon + \kappa\dot{H} - 3\kappa H^2 +
2\kappa H\f{\ddot{\phi}}{\dot{\phi}} \r) + p_{\rm tot} + \rho_{\rm tot}  \r].  \label{lala}
\ee
or
\be
\dot{H} =  \f{-4\pi G}{(1 + 4 \pi G \k \dot{\phi}^2)}       \l[ \dot{\phi}^2 \varepsilon + p_{\rm tot} + \rho_{\rm tot}
 - 3 \kappa \dot{\phi}^2 H^2 +
2\kappa H\ddot{\phi}\dot{\phi}        \r].  \label{lala2}
\ee
The factor ${G}/{(1 + 4 \pi G \k \dot{\phi}^2)}$ can not be regarded as $G_{\rm eff}$ since it is neither derived from the Einstein-frame Lagrangian nor the Friedmann equation in standard form.
The Klein-Gordon equation describes conservation of scalar field energy density. The equation can be viewed as modification in $\ddot{\phi}$ (field acceleration) and $\dot{\phi}$ (field speed) or modification of the scalar potential-slope term \cite{Sushkov:2009},
\be
\l(\varepsilon  - 3\kappa  H^2 \r)  \ddot{\phi}
+ \l(3 \varepsilon H
- 6\kappa H \dot{H} - 9 \kappa H^3\r)\dot{\phi}
= -V_{\phi}\,,    \label{eqf3}
\ee
or it can be rearranged as 
\be
\ddot{\phi} + 3H\dot{\phi} = - \f{V_{\phi}}{\varepsilon - 3\kappa H^2} + \f{6\kappa
H\dot{H}\dot{\phi}}{\varepsilon - 3\kappa H^2},   \label{KGmod}
\ee
where $V_{\phi} \equiv \d V/ \d \phi$. The field derivative of the effective potential is defined as
\be
V_{\rm eff, \phi}  =   \f{V_{\phi} - 6\kappa
H\dot{H}\dot{\phi}}{\varepsilon - 3\kappa H^2}\,,    \label{Veffrr}
\ee
which is a function expressed with choice of five variables $\phi, \dot{\phi}, \ddot{\phi}, H, \dot{H}$ of the system. Since there are three equations relating these variables, therefore there are only two degrees of freedom.
The energy density of barotropic and holographic dark energy are conserved separately as
\bea
   \dot{\rho}_{\Lambda} + 3H (\rho_{\Lambda} + p_{\Lambda}) & = & 0 \,, \label{ajaj2} \\
    \dot{\rho}_{\rm m} + 3H (\rho_{\rm m} + p_{\rm m}) & = & 0  \,,
\eea
where $\rho_{\Lambda}, p_{\Lambda}$ are density and pressure of the holographic dark energy contribution. The interaction case is to considered energy transfer between
 dark energy and matter in form of interaction term, $Q \propto H \rho_{\Lambda}$, i.e.    $ \dot{\rho}_{\Lambda} + 3H (\rho_{\Lambda} + p_{\Lambda})  =  Q $ and
$ \dot{\rho}_{\Lambda} + 3H (\rho_{\Lambda} + p_{\Lambda})  =  -Q $ such as in \cite{Wang:2007ak} and \cite{Jamil:2010xq}. Energy transfer from holographic dark energy to barotropic matter, ($Q > 0$)  would increase value of effective equation of state, resulting in less acceleration. The opposite sign of $Q$ gives opposite result. We do not consider the interaction case in this work.

\section{Holographic dark energy with NMDC gravity}
We shall apply concept of NMDC gravity to the holographic dark energy here.
Using $H^{-1}$ as holographic cutoff scale in equation (\ref{1}) and realizing that the gravitational constant is modified with the NMDC effect (equation (\ref{Geff})), hence
\be
\rho_{\Lambda}  =  \f{3 c^2 H^2}{8 \pi G_{\rm eff} }  =   \f{3 c^2}{8 \pi G} (1 + 12 \pi G \k \dot{\phi}^2) H^2\,.   \label{ajaj}
\ee
Hence the holographic modification tunes the NMDC density with the evolution of $H^2$, giving possibility of NMDC model to be viable.  The Friedmann equation (\ref{5})  now reads,
\be
H^2(1 - c^2)  =  \f{8\pi}{3}G_{\rm eff} \left(\frac{\varepsilon }{2}\dot{\phi}^2 + V  +  \rho_{\rm m}  \right)\,.
\ee
We consider late universe with dark energy domination evolving under the scalar potential
\be
V(\phi) = \f{1}{2}m^2 \phi^2\,.
\ee
 We choose this analytic potential because it is considered as free potential in  generalized Lagrangian that  includes free potential, free kinetic, non-minimal coupling (NMC) and NMDC terms.  The $\phi^2$-dependency makes it comparable to the NMC, $R \phi^2$  term which plays a role as NMC potential  in \cite{Setare:2008pc}.
The barotropic density is negligible and the solutions are considered in form of
\bea
  a  & = &  a_0 e^{r t}\,,   \\
  \phi  & = &  \phi_0 e^{s t}\,,
\eea
or one can use  $a = a_0 e^{r x}$  and  $\phi = \phi_0 e^{s x}$ where the e-folding number, $N \equiv r x$. Using either forms of the solution results in the same modified Friedmann equation (negligible $\rho_{\rm m}$) and Klein-Gordon equation,
\be
\varepsilon s^2    =  -m^2\;\;\;\;\;\;\;\text{and}\;\;\;\;\;\;\;   \\
 s^2  + 3 r s + \f{m^2}{\varepsilon - 3 \kappa r^2 }     =    0\,,         \label{qpqp}
 \ee
 where the constant $c$ is set to 1.  This gives a relation,
 \be
 s  \:= \: \f{\varepsilon - 3 \kappa r^2}{\kappa r}\,,      \label{qpqp2}
 \ee
where conditions $\k \neq 0$, $\;r \neq 0$ and $s \neq 0$ are imposed for those cases that apply this relation. This scale factor function and the scalar field solution are hence  solutions of the system\footnote{Formal approach, i.e. dynamical stability analysis, is needed, in future work, to check other possible fixed points and their stability solutions.}.
If without scalar potential $V = 0$, i.e. $ m = 0 = s$, we can not use the equation (\ref{qpqp2}). Instead, we must use the equation (\ref{qpqp}). Then one can see that $r$ can take any real values. In case of existing non-negligible barotropic density $\rho_{\rm m}$ in the Friedmann equation,
\be
\f{\varepsilon}{2} \phi_0^2 s^2 e^{2 s t}  +  \f{1}{2} m^2 \phi_0^2 e^{2st}  + \f{\rho_{{\rm m},0}}{a_0^{3(1+w_{\rm m})} e^{3(1+w_{\rm m})rt}}  \;=\;  0\,.  \label{FR22}
\ee
We can see that $m=0$ (that is $V=0$) does not imply $s=0$ but implying phantom field, $\varepsilon < 0$.
When setting $\varepsilon = 0$, i.e. removing free kinetic term, the Lagrangian is left with purely NMDC kinetic effect (absorbed into $G_{\rm eff}$), scalar potential term and barotropic matter term. This results in either $m^2 < 0$ or $\rho_{{\rm m},0} < 0$ which are not realistic. If we consider all three terms together, it also implies  phantom field, $\varepsilon < 0$. For simplification in this work, we consider that the barotropic density is negligible and we consider the relation (\ref{qpqp2}) in three cases of the $\varepsilon$ value.

\section{Holographic dark energy with purely NMDC kinetic term ($\varepsilon = 0$ )}
\begin{figure}[t]  \begin{center}
\includegraphics[width=6.0cm,height=3.8cm,angle=0]{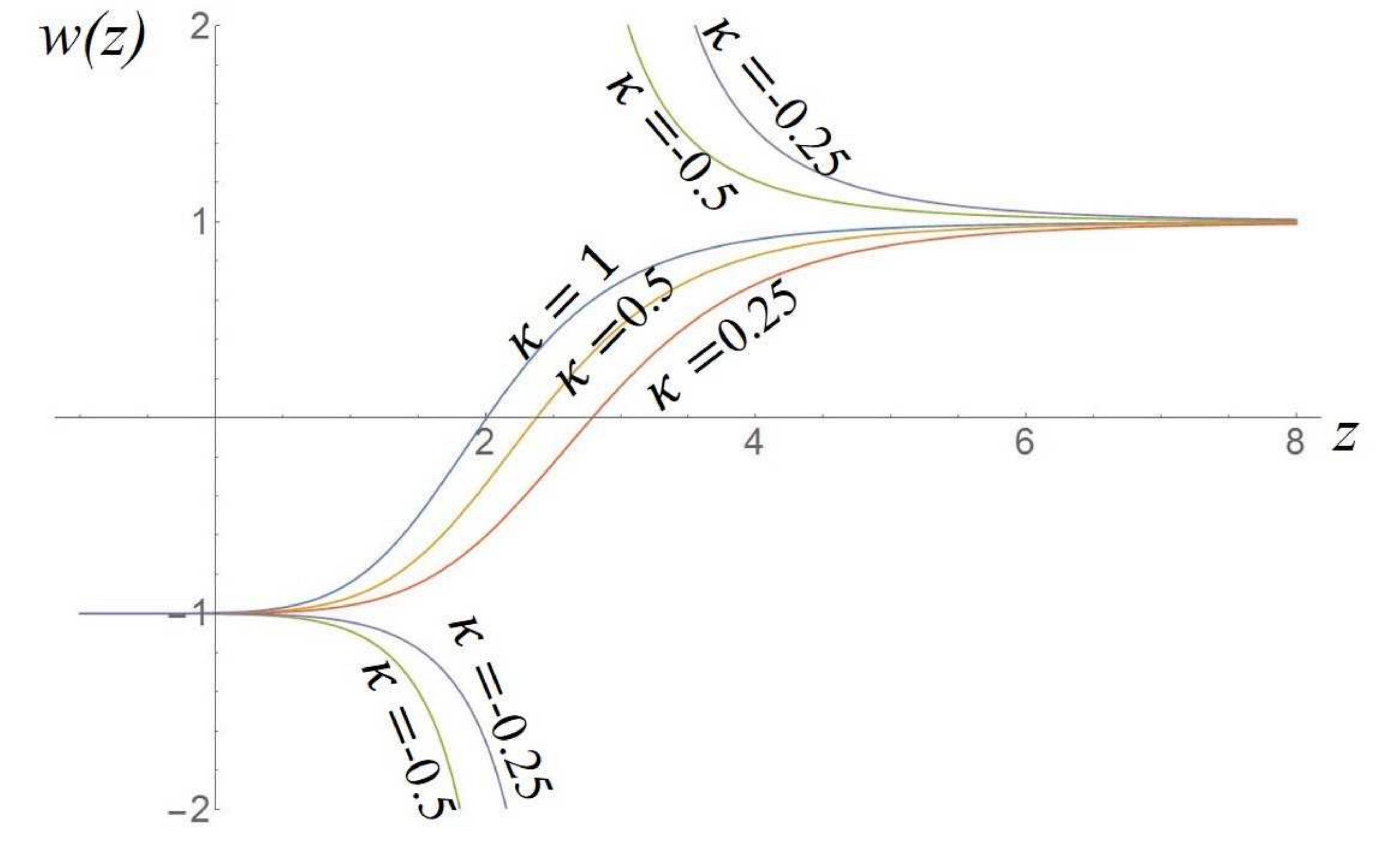}  \end{center}
\caption{Equation of state for the holographic dark energy with purely NMDC kinetic term is plotted versus $z$ for $\kappa = -0.5, -0.25, 0.25, 0.5, 1.0$ and $r = 0.01$. Divergencies in the case $\k < 0$ are predicted with singularity in equation (\ref{wpureNMDC}).  \label{fig1}} \begin{center}
\includegraphics[width=6.0cm,height=3.8cm,angle=0]{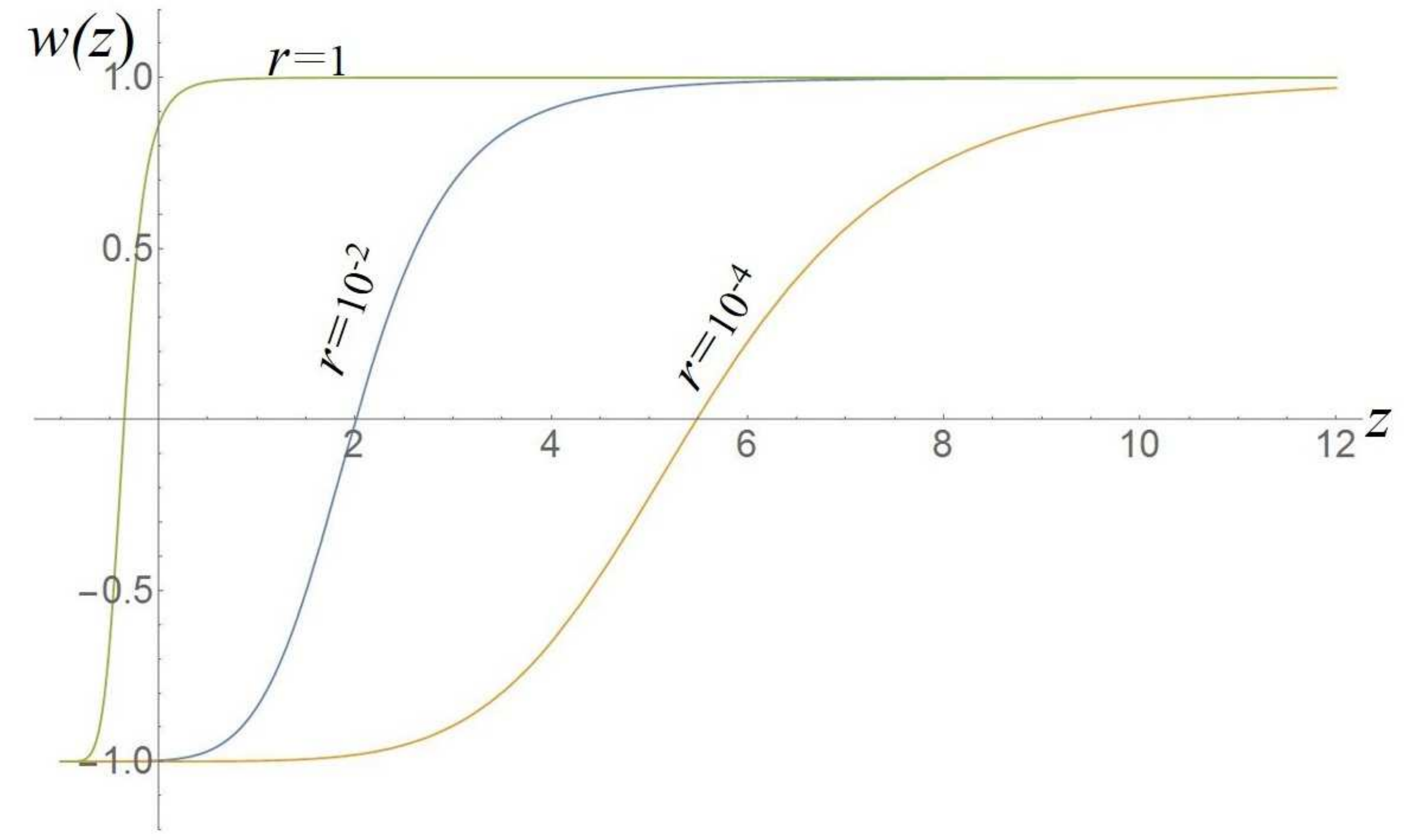}  \end{center}
\caption{Equation of state for the holographic dark energy with purely NMDC kinetic term plotted against $z$. Keeping $\kappa = 1.0$ and varying $r = 1, 0.01, 0.0001$, less $r$ makes $w(z)$ approaching at higher redshift.
If  $r$ is as less as  $H_0 \sim \sqrt{\Lambda/3} \sim 10^{-121}$,  as a result $w(z)$ must have approached $-1$ long time ago.
   \label{fig2}}  \end{figure}

The first case is to consider purely NMDC kinetic theory. That is the gravitational coupling to the field derivative is the only kinetic term, i.e. we let $\varepsilon = 0$. Therefore, from equation (\ref{qpqp2}),  $s = -3r$ (with negligible $\rho_{\rm m}$ and $s \neq 0$). The solutions are
\be
 a   =   a_0 e^{r t}\,, \;\;\;\; \phi   =   \phi_0 e^{-3r t}\,.
\ee
Using these results in equations (\ref{ajaj2}) and (\ref{ajaj}), the energy density of the holographic dark energy is (keeping $c$ here for completeness)
\be
\rho_{\Lambda}  =    \f{3 c^2}{8\pi G}  \l( 1 + 108 \pi G \k \phi_0^2 r^2 e^{-6rt}   \r) r^2\,,
\ee
and the pressure is
\be
p_{\Lambda}  =    - \f{3 c^2}{8\pi G}  \l( 1 - 108 \pi G \k \phi_0^2 r^2 e^{-6rt}   \r) r^2\,.
\ee
This gives the equations of state,
\be
w_{\Lambda} =  \f{ -1 + 108 \pi G \k \phi_0^2 r^2 e^{-6rt} }{ 1 + 108 \pi G \k \phi_0^2 r^2 e^{-6rt}}\,.
\ee
Converting $t$ into redshift $z$ with relation $1+z = a_0/a$, we have, \be -rt = \ln(1+z). \ee
In this convention, at present $t = t_0 = 0$ as $z =0$ and $t \rightarrow -\infty$ as $z \rightarrow \infty$ in the past.  Therefore
\be
w_{\Lambda} =  \f{ -1 + 108 \pi G \k \phi_0^2 r^2 (1+z)^6  }{ 1 + 108 \pi G \k \phi_0^2 r^2  (1+z)^6}\,.  \label{wpureNMDC}
\ee
The present Hubble parameter $H_0$ is considered as $r$, we think of sub-Planckian values of $\phi_0$ and of the coupling constant $\k$. Various values of these parameters are used and the evolution of $w_{\Lambda}(z)$ is presented in Fig. \ref{fig1}.   This case is interesting since it renders $w_{\Lambda} \longrightarrow -1$ at late time. However $\k < 0$ case is not favored because its $w_{\Lambda}$ equations either goes out of the range $[-1,1]$ or comes from outside the range $[-1,1]$ for all evolutions.   All numeric values used here are in Planck unit, i.e.  we set $8\pi G = 1, \phi_0 =1 $.
Considering $r \sim H_0 \sim \sqrt{\Lambda/3}$ and $\Lambda/M_{\rm P}^2 \sim 10^{-121}$ (the reduced Planck mass $M_{\rm P} = (8 \pi G)^{-1/2}$.), hence $r \sim 10^{-60}$. The less $r$ brings $w_{\Lambda}$ to $-1$ earlier as shown in Fig. \ref{fig2} in which two different values of $r$ are used while fixing other variables. At present $z=0$, singularity in $w_{\Lambda}$ occurs if $\k = \k_{{\rm s}, z=0} = - (108 \pi G \phi_0^2 r )^{-1} $ which is $\k_{{\rm s}, z=0} =  -7.4 \times 10^{58}$ in Planck unit. Singularity in $\k$ is negative. Since we favor positive $\k$ case, it is not a problem that $\k_{{\rm s}, z=0} < 0$.

\section{Holographic dark energy with both NMDC and free kinetic terms}
For $\varepsilon = \pm 1$, the relations (\ref{qpqp}) and (\ref{qpqp2}) allow us to express $r$ in terms of $m$ and $\kappa$,
\be
r   =  \pm    \f{\sqrt{\f{(6 - m^2 \kappa)}{\varepsilon}  \pm   m \sqrt{\kappa (m^2 \kappa - 12)}   }}{\sqrt{18 \kappa}}\,,   \label{29}
\ee where the condition $s \neq 0, r \neq 0, \k \neq 0$ must hold.
Energy density and pressure of the holographic dark energy are
\be
\rho_{\Lambda}  =    \f{3 c^2}{8\pi G}  \l( 1 + 12 \pi G \k \phi_0^2 s^2 e^{2 s t}   \r) r^2\,,
\ee
and
\be
p_{\Lambda}  =    - \f{3 c^2}{8\pi G}  \l( r^2 +  8 \pi G \k \phi_0^2 s^3 e^{2st} r  + 12 \pi G \kappa \phi_0^2 s^2 e^{2 s t} r^2   \r)\,.
\ee
The equation of state is hence
\be
w_{\Lambda}   =   - \f{\l(r +  8 \pi G \k \phi_0^2 s^3 e^{2st} +  12 \pi G \k \phi_0^2 s^2 e^{2 s t} r \r)}{ r \l( 1 + 12 \pi G \k \phi_0^2 s^2 e^{2st}  \r)}\,,
\ee
or, as function of redshift,
\be
w_{\Lambda}   =    - \f{\l[ r + 8 \pi G \k \phi_0^2  s^3  (1+z)^{-\f{2s}{r}}   +  12 \pi G \k \phi_0^2 s^2 r  (1+z)^{-\f{2s}{r}} \r]}{ r \l[ 1 + 12 \pi G \k \phi_0^2 s^2 (1+z)^{-\f{2s}{r}}   \r]   },
\ee
where the relation $-rt = \ln(1+z)$ is expressed as $e^{2st} = (1+z)^{-\f{2s}{r}}$. In equation (\ref{qpqp}), $ \varepsilon s^2 = -m^2$,
the value of $s$ is real if $ \varepsilon = -1 $, i.e. the kinetic term is phantom. On the other hand, if $\varepsilon = 1$, $s$ is imaginary.

\begin{figure}[h!]
\begin{center}
\includegraphics[width=6.2cm,height=3.5cm,angle=0]{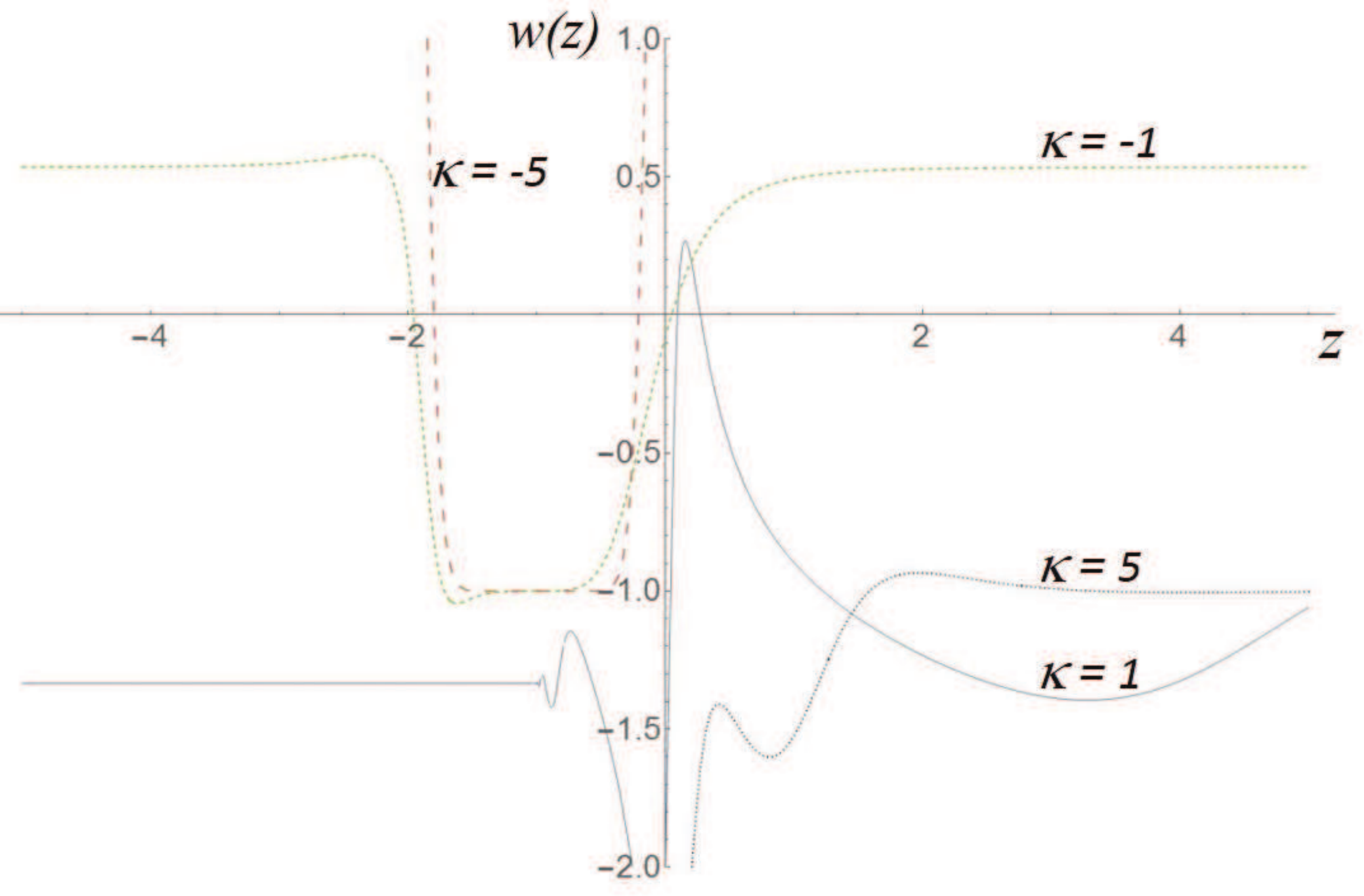}
\end{center}
\caption{Real part of the equation of state of the {\bf case 1 and case 6} of canonical kinetic term case for $m = 1$ and $\kappa = -5, -1, 1, 5$. \label{fig7}}
\begin{center}
\includegraphics[width=6.2cm,height=3.5cm,angle=0]{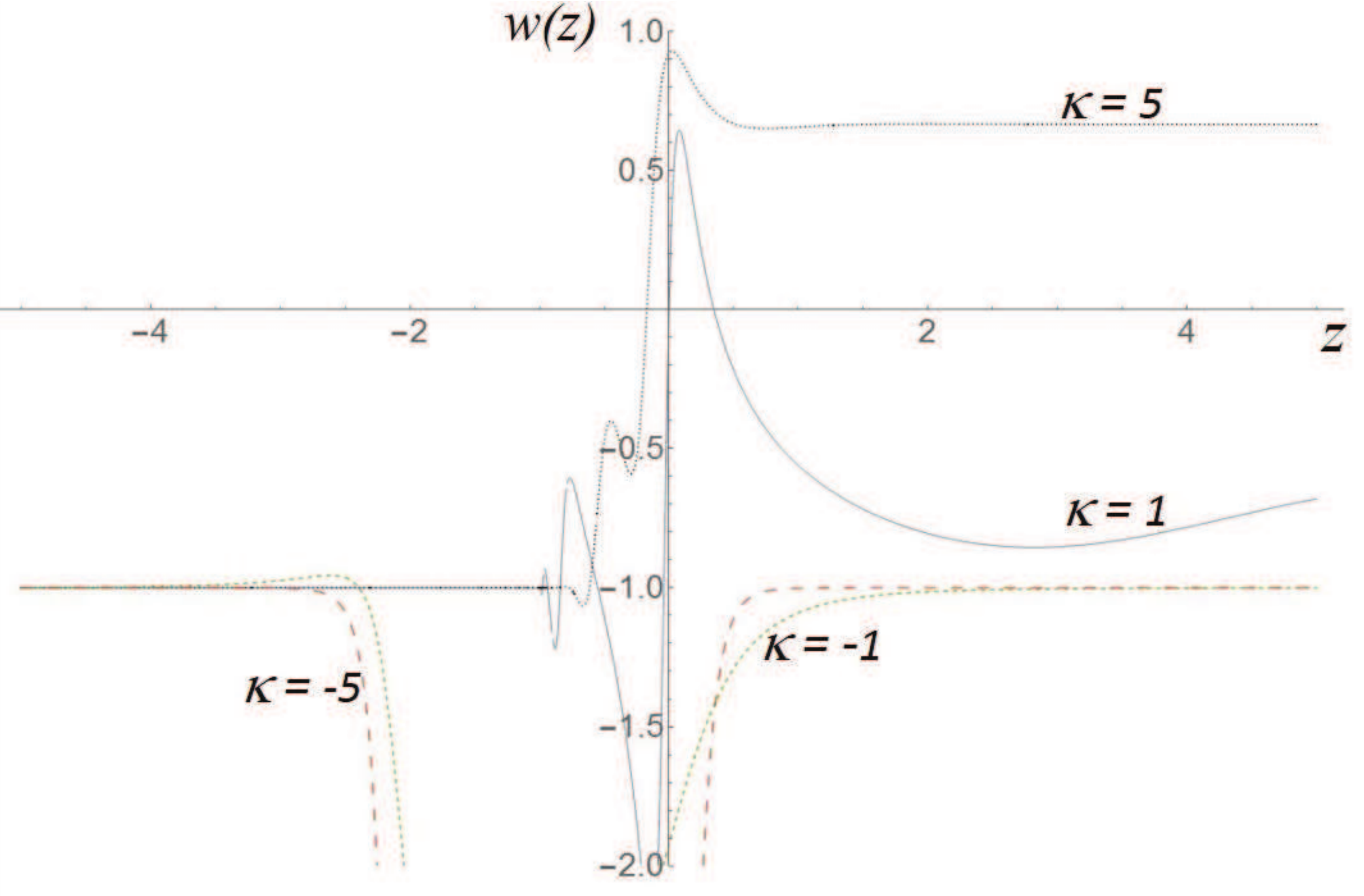}  \end{center}
\caption{Real part of the equation of state of the {\bf case 2 and case 5} of canonical kinetic term case for $m = 1$ and $\kappa = -5, -1, 1, 5$.  \label{fig8}}
\begin{center}
\includegraphics[width=6.2cm,height=3.5cm,angle=0]{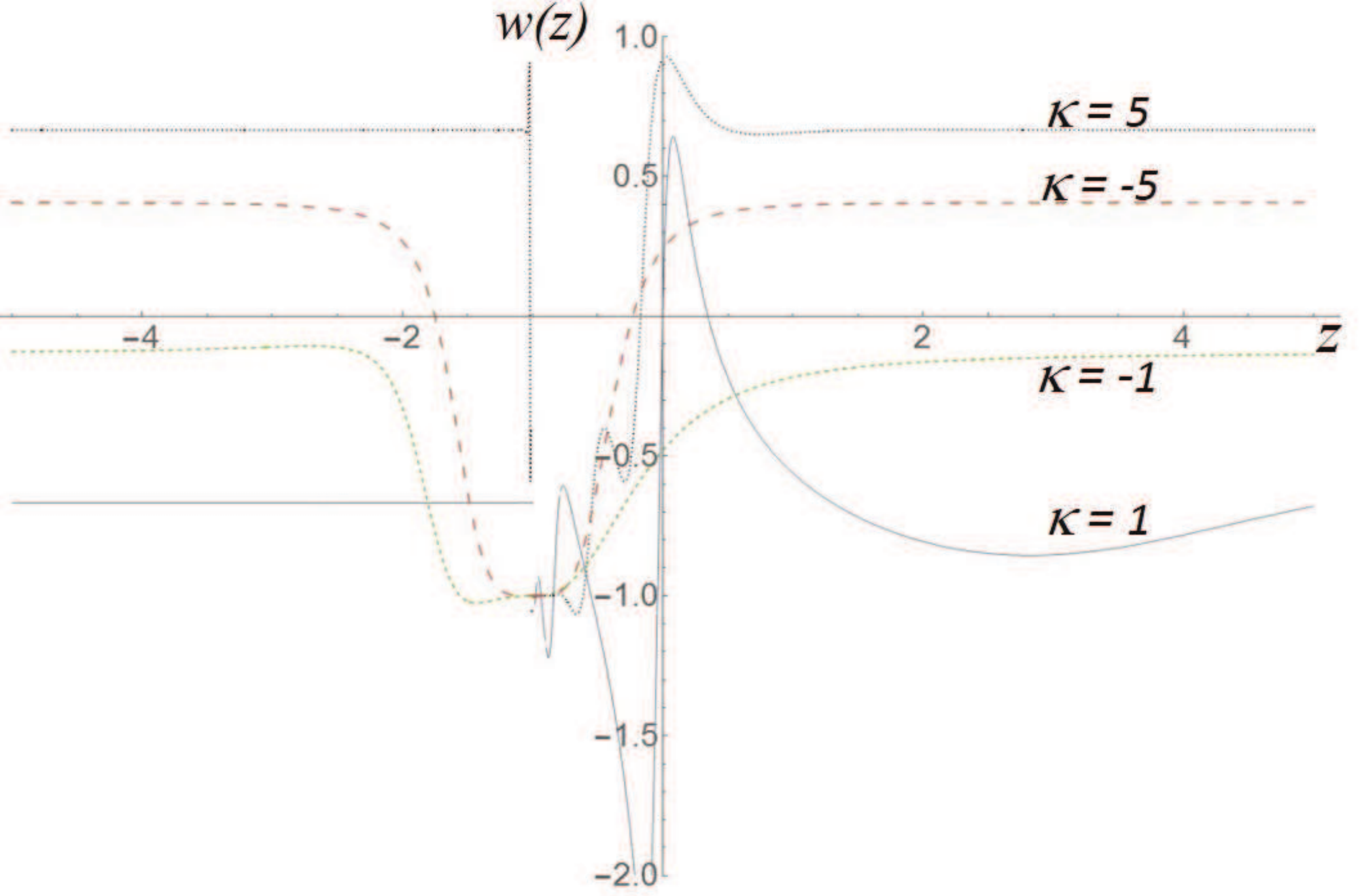}  \end{center}
\caption{Real part of the  equation of state of the {\bf case 3 and case 8} of canonical kinetic term case for $m = 1$ and $\kappa = -5, -1, 1, 5$.  \label{fig9}}
\begin{center}
\includegraphics[width=6.2cm,height=3.5cm,angle=0]{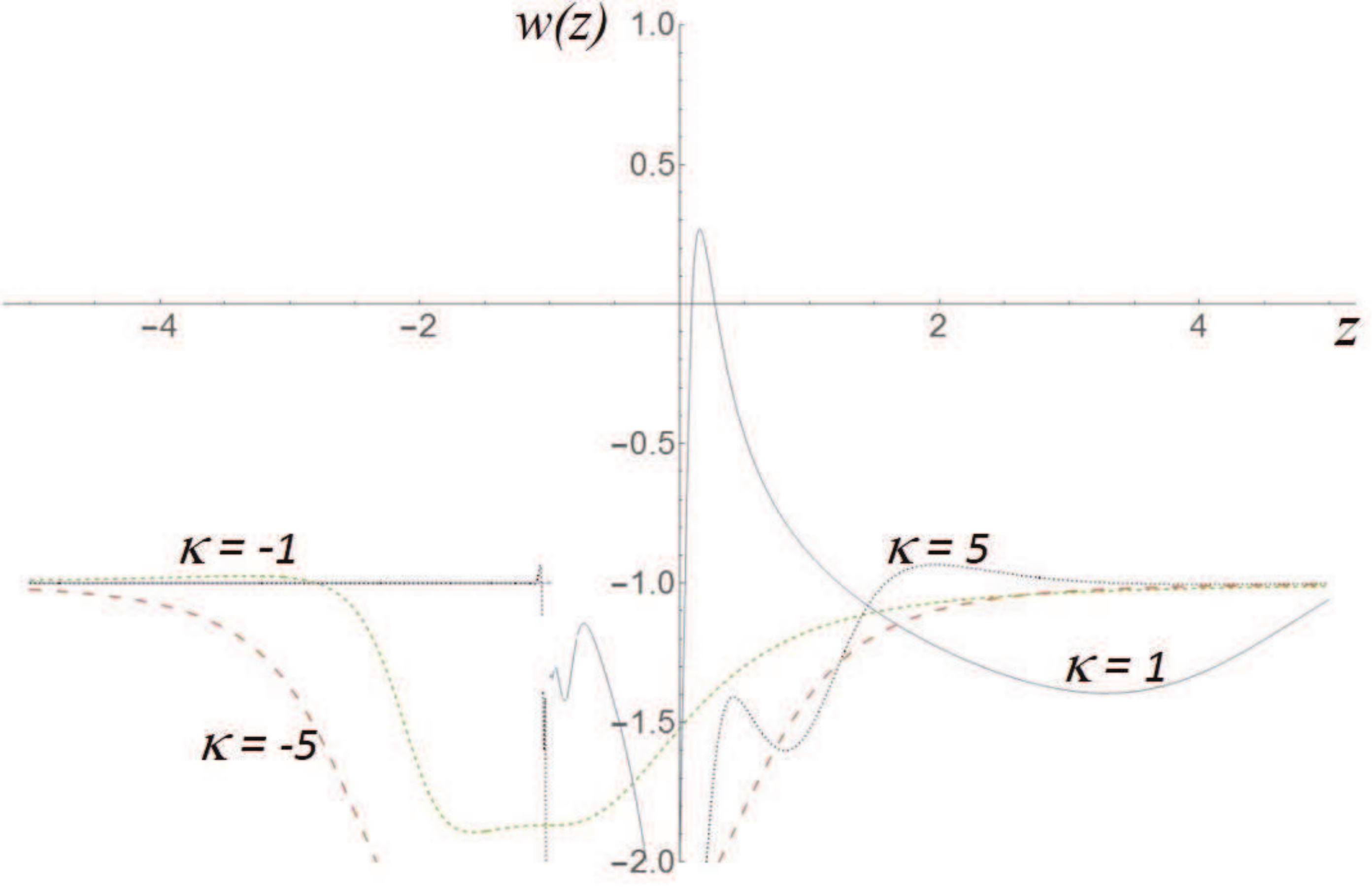}  \end{center}
\caption{Real part of the equation of state of the {\bf case 4 and case 7} of canonical kinetic term case for $m = 1$ and $\kappa = -5, -1, 1, 5$.  \label{fig10}}
\end{figure}

\subsection{Canonical kinetic term  ($\varepsilon = 1$)}   \label{ohmygod}
With free canonical kinetic term $\varepsilon = 1$, $s^2 = -m^2$ hence $s  =  \pm i m  $ which is imaginary and $s^3 = \mp i m^3 $.
The equation of state as function of time is
\bea
w_{\Lambda} = - \f{1}{C}\l( A + i B \r)\,,   \label{koko}
\eea
where $A, B$ are real and imaginary parts,
\bea
& & A \equiv  1 + \no \\
& & 8\pi G \k \phi_0^2 m^2 \l[ -3 \cos(2 m t)
 + \f{m}{r}\sin(2mt) + 18 \pi G \k \phi_0^2 m^2 \r]\,, \no\\ \label{jojojo} \\
& & B \equiv  8\pi G \k \phi_0^2 \f{m^3}{r}    \l[ \pm 12 \pi G \k \phi_0^2 m^2 \mp  \cos(2 m t) \r], \label{42}
\eea
and
\be
C \equiv 1 + 24 \pi G \k \phi_0^2 m^2 \l[ -\cos(2mt)   +  6 \pi G \k \phi_0^2 m^2\r].
\ee
The $\pm$ and $\mp$ signs of equation (\ref{42}) correspond to the sign of $s = \pm i m $ consequently.
It should be noted that complex value of $w_{\Lambda}$ does not imply that NMDC density or free scalar kinetic term are complex, but the complex value arises from holographic modification of the Friedmann equation, $\varepsilon s^2 = - m^2 $ (\ref{qpqp}) and
in deriving $p_{\Lambda}$ from the fluid equation (\ref{ajaj2}).  The complex value is indeed of the $\rho_{\Lambda}, p_{\Lambda}$,  not of the scalar field density nor pressure terms.
There is a singularity in $w_{\Lambda}$ when
\be
t_{\rm s} = \f{1}{2m} \arccos\l[\f{1+ 144 \pi^2 G^2 \k^2 \phi_0^4 m^4 }{24 \pi G \k \phi_0^2 m^2}\r], \label{yhfi}
\ee
for a fixed $\k \neq 0$. If we need to know the range of $\k$ that can give singularity at present epoch, we use $t=t_0 = 0$
at present time and assume $m = 1$ in equation (\ref{yhfi}). The result is $\k_{{\rm s},t_0} = 2/3$.
As argument of $\arccos$ function in equation (\ref{yhfi}) must be in the range $[-1,1]$, this puts the limits to $\k_{\rm s}$ to be in the range $[-2/3, 2/3]$ which limits $t_{\rm s}$ into $[0, \pi]$ in Planck unit.
If considering uncarefully, we might think that present value $w_{\Lambda, t_0}$ (as a function $\k$ in Planck unit, setting $8 \pi G \equiv 1, m = 1, \phi_0 =1$) can be found from the real part $-A/C$ in equation (\ref{koko}) as $w_{\Lambda, t_0} = [-1+3\k-(9/4)\k^2]/[1-3\k+(9/4)k^2] = -1$. However this is not correct. The coefficient $r$ in equation (\ref{jojojo}) could take imaginary value and need to be taken into account. Hence considering equation (\ref{29})  in our $\varepsilon = 1$ context,
\bea
r  &  = &  \pm     \f{\sqrt{ 6  - m^2 \k \l(1  \mp    \sqrt{1  - \f{12}{m^2 \kappa}}\r)}}{\sqrt{18 \kappa}}\,.
\eea
The $\pm$ and $\mp$ signs in the expression of $r$ come from solving quadratic equation. Here $r$ always  has  complex value.  Using relation $t = - r^{-1} \ln(1+z)$, we plot real part of $w_{\Lambda}(z)$ in all possible eight cases.  To simplify, we define
$D \equiv  \sqrt{1 - 12/(m^2 \k)}$, $E \equiv 8\pi G \k \phi_0^2 m^3/r    $ and $F \equiv 12 \pi G \k \phi_0^2 m^2    $. Conditions of the eight cases are,
\begin{itemize}
  \item  {\bf case 1}: \\
  $
r    =   {\sqrt{ 6  - m^2 \k \l(1  -    D  \r)}}/{\sqrt{18 \kappa}}\,,\\
B  =   E    \l[  F   -  \cos(2 m t) \r],\;s =  i m
$
   \item  {\bf case 2}: \\
$
r    =     {\sqrt{ 6  - m^2 \k \l(1  -    D  \r)}}/{\sqrt{18 \kappa}}\,,\\
B  =   E    \l[ - F   +  \cos(2 m t) \r],\; s = -  i m
$
    \item  {\bf case 3}: \\
$
r    =     {\sqrt{ 6  - m^2 \k \l(1  +    D  \r)}}/{\sqrt{18 \kappa}}\,,\\
B  =   E    \l[  F   -  \cos(2 m t) \r],\;s =  i m
$
     \item  {\bf case 4}: \\
$
r    =     {\sqrt{ 6  - m^2 \k \l(1  +    D  \r)}}/{\sqrt{18 \kappa}}\,,\\
B  =   E    \l[ - F   +  \cos(2 m t) \r],\; s = -  i m
$
      \item  {\bf case 5}: \\
$
r    =    - {\sqrt{ 6  - m^2 \k \l(1  -    D  \r)}}/{\sqrt{18 \kappa}}\,, \\
B  =   E    \l[  F   -  \cos(2 m t) \r],\;s =  i m
$
       \item  {\bf case 6}: \\
$
r    =     - {\sqrt{ 6  - m^2 \k \l(1  -    D  \r)}}/{\sqrt{18 \kappa}}\,,  \\
B  =   E    \l[ - F   +  \cos(2 m t) \r],\; s = -  i m
$
        \item  {\bf case 7}: \\
 $
r    =     -{\sqrt{ 6  - m^2 \k \l(1  +    D  \r)}}/{\sqrt{18 \kappa}}\,,  \\
B =   E    \l[  F   -  \cos(2 m t) \r],\;s =  i m
$
         \item  {\bf case 8}: \\
          $
r   =     -{\sqrt{ 6  - m^2 \k \l(1  +    D  \r)}}/{\sqrt{18 \kappa}}\,,\\
B  =   E    \l[ - F   +  \cos(2 m t) \r],\; s = - i m  \,.
$
\end{itemize}
Real parts of the $ w_{\Lambda}(z) $ for these cases are plotted. Real parts of some cases are the same. These are (case 1 and case 6: Fig. \ref{fig7}),  (case 2 and case 5: Fig. \ref{fig8}), (case 3 and case 8: Fig. \ref{fig9}) and (case 4 and case 7: Fig. \ref{fig10}).  Moreover, when $\k > 0$, real parts of
$w_{\Lambda}(z)$ in Fig. \ref{fig7} and Fig. \ref{fig10} are the same and the real parts of $w_{\Lambda}(z)$ in Fig. \ref{fig8} and Fig. \ref{fig9} are also the same. This is because the distinct of each case appears in the imaginary parts.  Cases 2  and 5 (Fig. \ref{fig8}) and cases 4 and 7  (Fig. \ref{fig10}) have $-1$  as late time value of $w_{\Lambda}$ while the rests do not. Considering realistic character that $w_{\Lambda}$ should be about $-1$ at present, and past evolution should not have $w_{\Lambda}< -1$ hence
only reasonable cases are the portraits with $\k > 0$ in Fig. \ref{fig8}. These are of case 2 and case 5. Focusing on case 2, we take positive root of $r$ with $s = - i m$ while for case 5 we take negative root of $r$ with $s = im$. These two cases result in the same real part of $w_{\Lambda}$ but we need the expansion to be de-Sitter and positive root of $r$ is preferred which matches only the case 2.

\subsection{Phantom kinetic term  ($\varepsilon = -1$)}  \label{sec_phantom}
\begin{figure}[t]  \begin{center}
\includegraphics[width=6cm,height=3.8cm,angle=0]{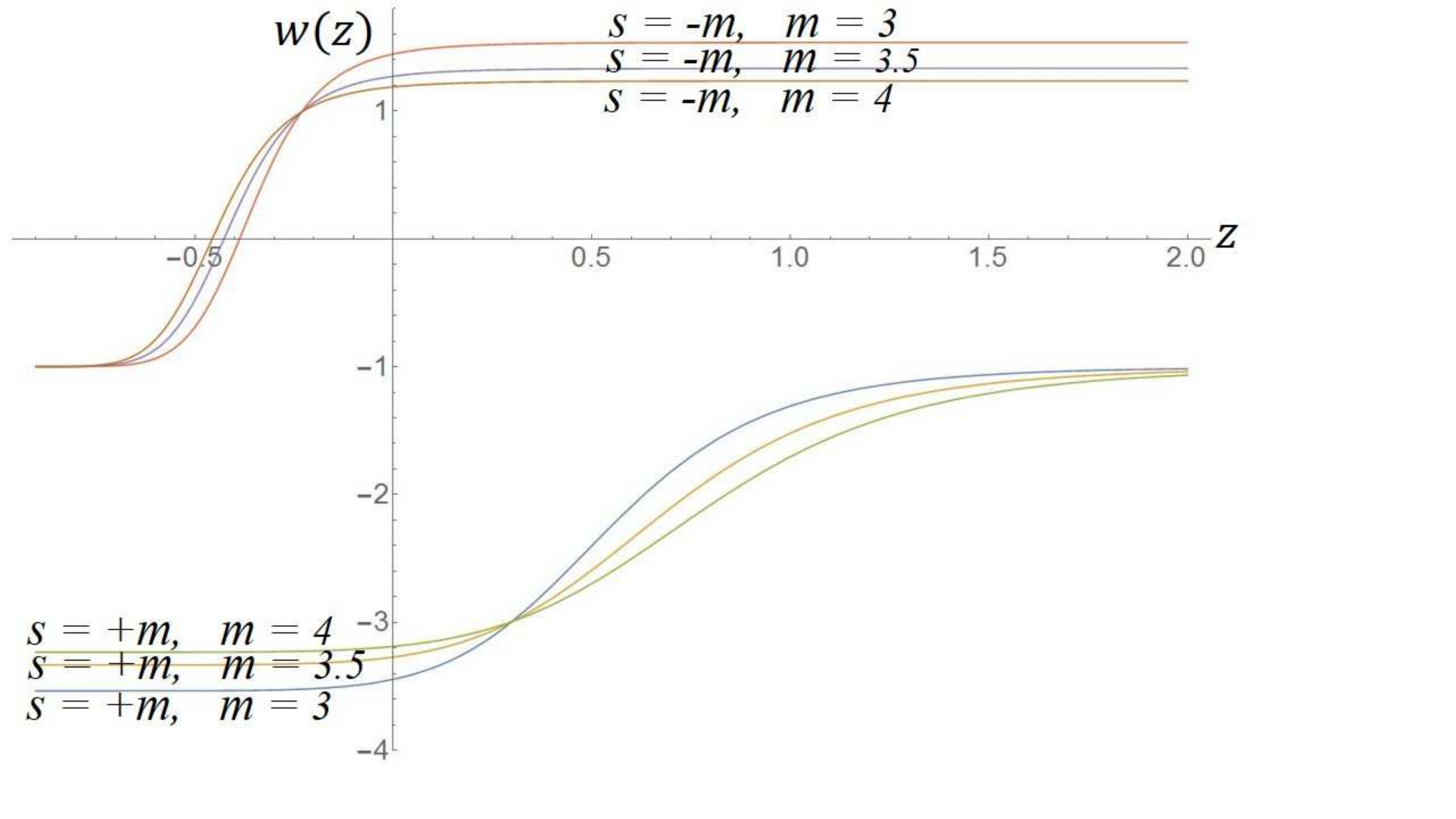}  \end{center}
\caption{Equation of state of the positive branch of the equation (\ref{branch1}) for  $\kappa = 2$ and $m = 3, 3.5, 4$. \label{fig3}}
\begin{center}
\includegraphics[width=6cm,height=3.8cm,angle=0]{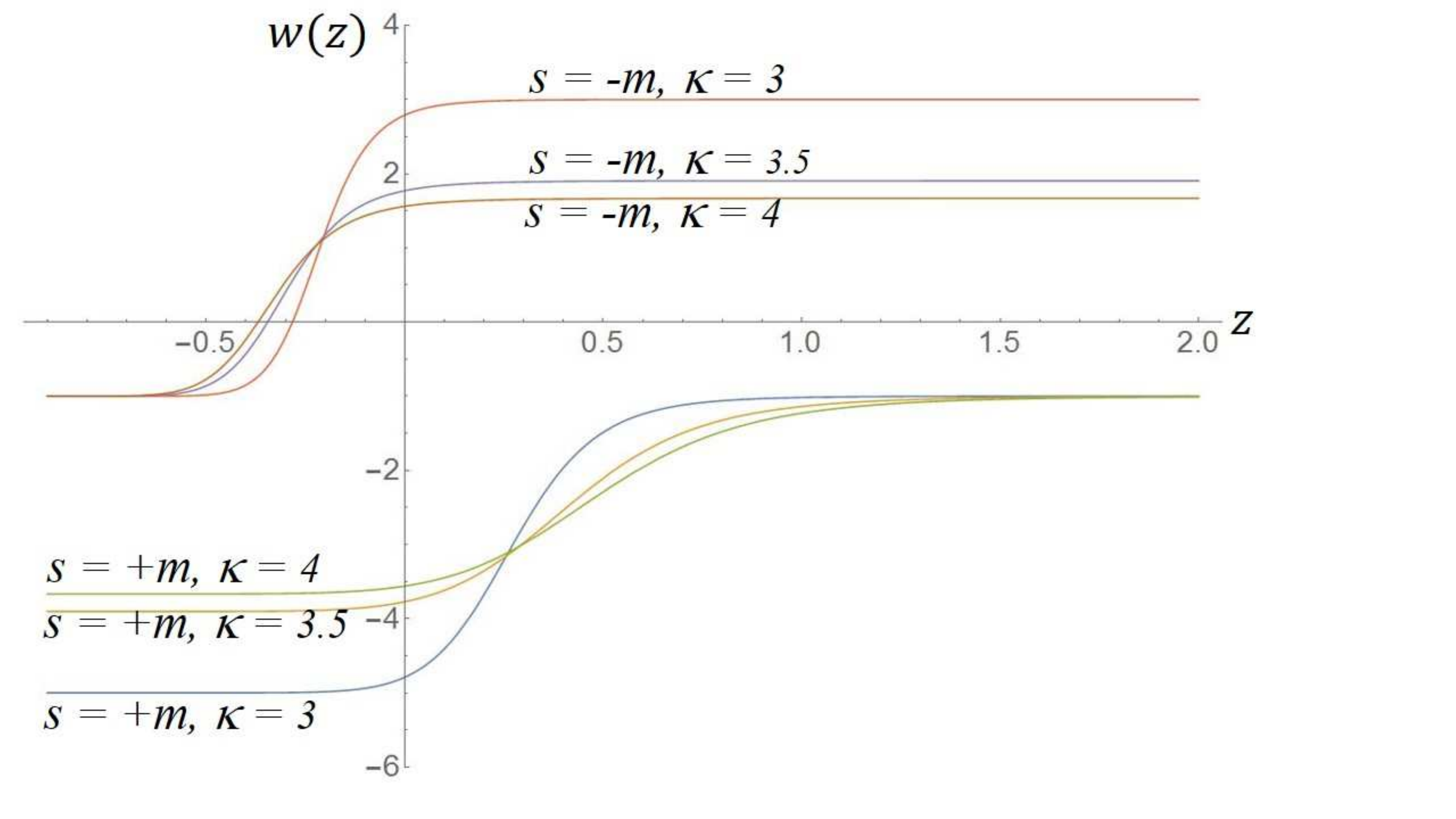}  \end{center}
\caption{Equation of state of the positive branch of the equation (\ref{branch1}) for $\kappa = 3, 3.5, 4$ and $m = 2$.  \label{fig4}}  \end{figure}

\begin{figure}[t]
\begin{center}
\includegraphics[width=6cm,height=3.8cm,angle=0]{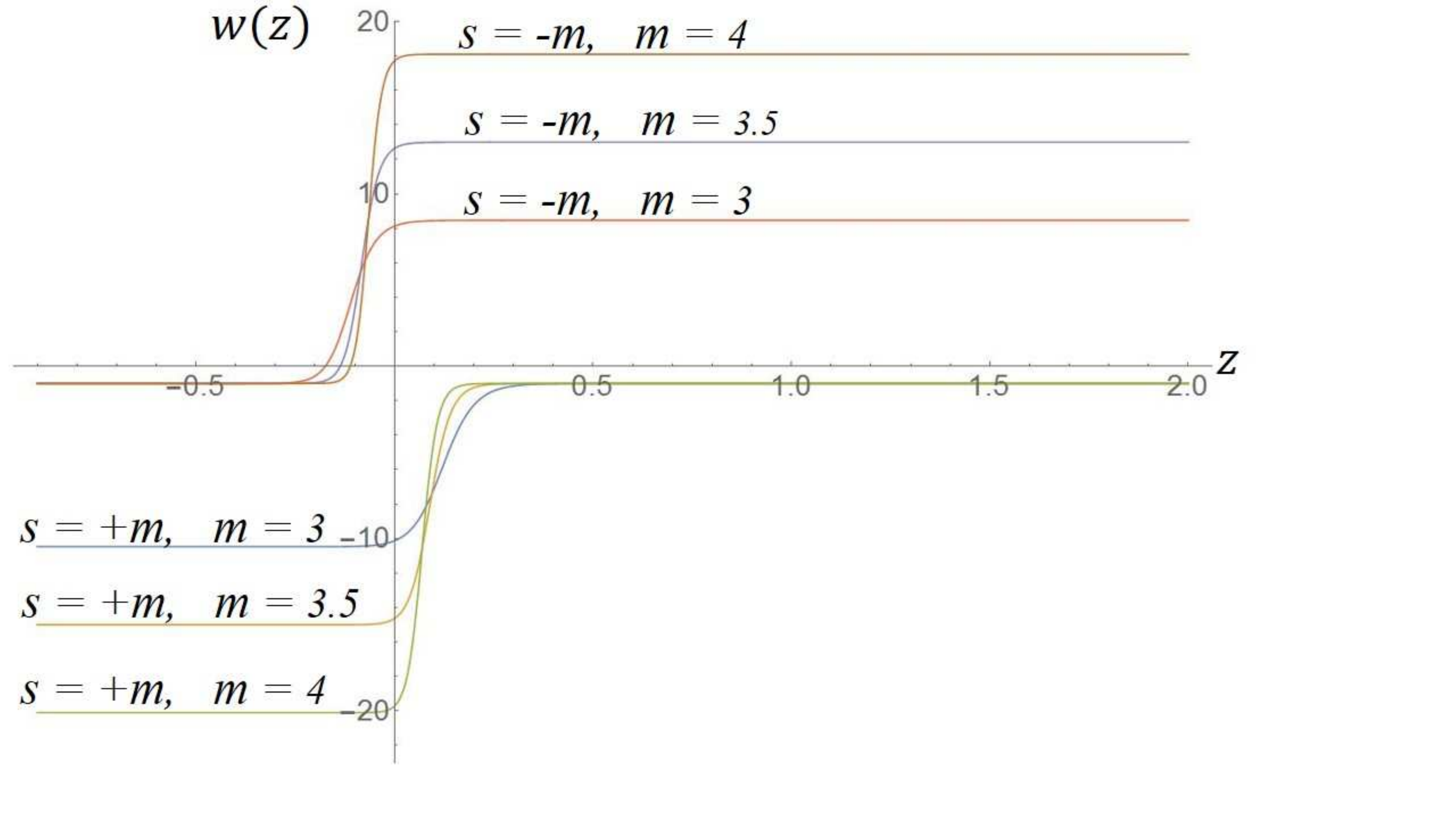}  \end{center}
\caption{Equation of state of the negative branch of the equation (\ref{branch1}) for  $\kappa = 2$ and $m = 3, 3.5, 4$. \label{fig5}}    \begin{center}
\includegraphics[width=6cm,height=3.8cm,angle=0]{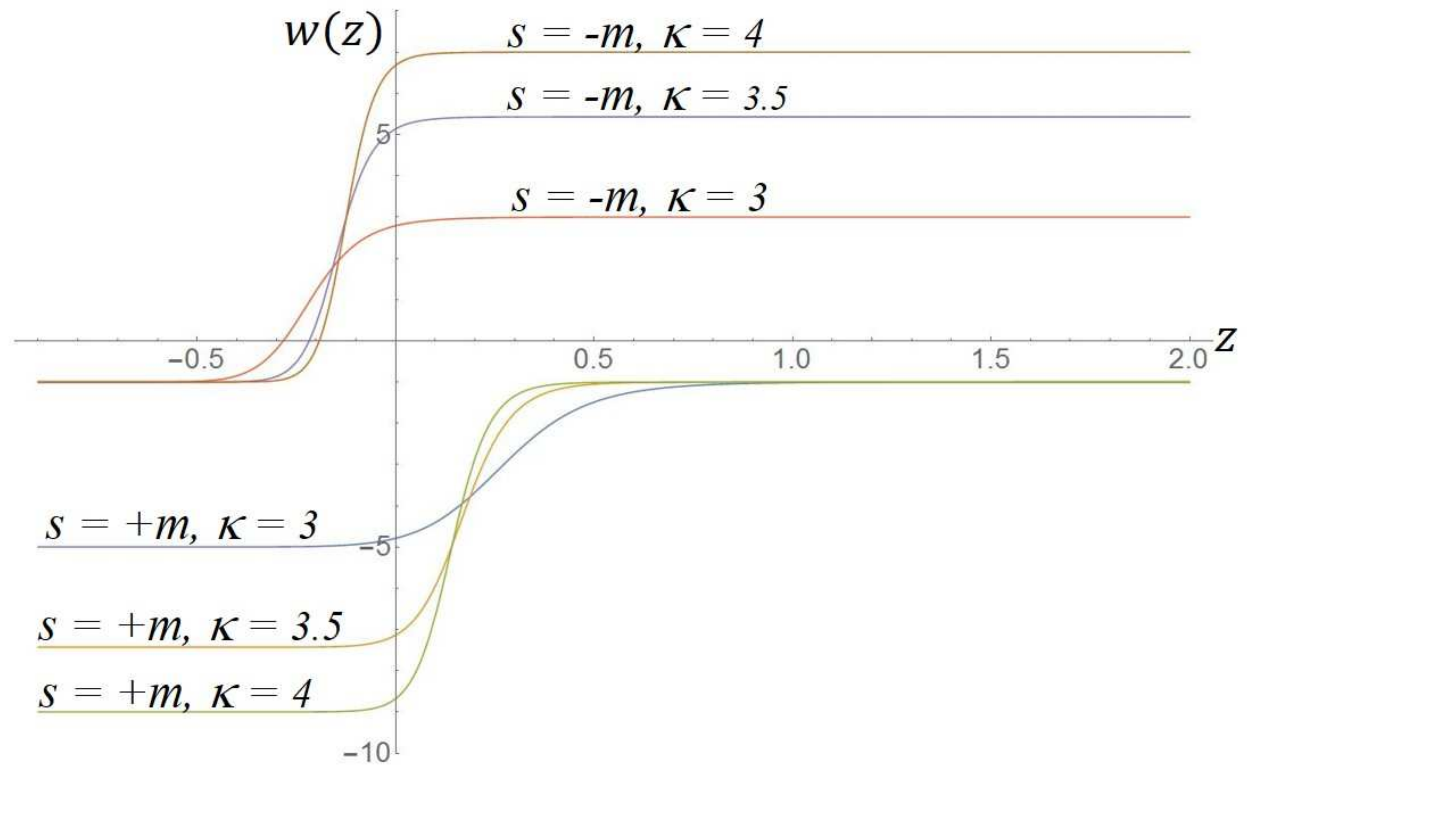}  \end{center}
\caption{Equation of state of the negative branch of the equation (\ref{branch1}) for $\kappa = 3, 3.5, 4$ and $m = 2$. \label{fig6}}  \end{figure}

Considering phantom kinetic term $\varepsilon = -1$, we have $s  =  \pm m $.  The equation of state is
\bea
&& w_{\Lambda}   =  \no \\
&&  - \f{\l[ r \pm 8 \pi G \k \phi_0^2 m^3  (1+z)^{\mp \f{2m}{r}}   +  12 \pi G \k \phi_0^2 m^2 r  (1+z)^{\mp \f{2m}{r}} \r]}{ r \l[ 1 + 12 \pi G \k \phi_0^2 m^2 (1+z)^{\mp  \f{2m}{r}}   \r]   }\,, \no \\
\eea
where the $\pm$ and $\mp$ sign correspond to $s = \pm m$ accordingly. The $\pm$ signs in the expression of $r$ (equation (\ref{29})) do not result from $s = \pm m$ but are positive and negative roots obtained in solving the equations of motion (\ref{5})  and  (\ref{eqf3}). Here we have
\bea
r  &  = &       \f{\sqrt{ -6  + m^2 \k \l(1  \pm    \sqrt{1  - \f{12}{m^2 \kappa}}\r)}}{\sqrt{18 \kappa}}\,.  \label{arrrr}
\eea
The scale factor is spatial expansion when $r$ is real and positive, that is $\k > 0$ and
\be
     \pm    \sqrt{1  - \f{12}{m^2 \kappa}}  \: >  \:   \f{6}{m^2 \k} - 1\,.  \label{branch1}
\ee
The right-hand side requires that
\be
m^2 \kappa \geq 12   \;\;\;\;\text{or}\;\;\;\; m \geq \f{2\sqrt{3}}{\sqrt{\k}}\,, \ee resulting that $\sqrt{1  - {12}/({m^2 \kappa})}$ falls into a range $[0, 1)$ for the positive branch of the left-hand side of equation (\ref{branch1}).
This also restricts the value of $({6}/{m^2 \k}) - 1$ to a range $(-1,-0.5]$.  The negative branch of the left-hand side is restricted to $(-1, 0]$. The negative branch corresponds to value of $({6}/{m^2 \k}) - 1$ to lie within $(-1,-0.5]$.
The situation requires both coupling $\k>0$ and the scalar mass $m$ to be in super-Planckian regime as we set $\phi_0 = 1$. Unless positive $\k$ nor $\k =0$, scalar mass is imaginary.
 Fig. \ref{fig3} and Fig. \ref{fig4} present the plots of $w_{\Lambda}(z)$ for the positive branch of equation (\ref{branch1}) and Fig. \ref{fig5} and Fig. \ref{fig6} presents the plots of $w_{\Lambda}(z)$ for the negative branch. Combination of the values of $m$ and $\k$ results in the value of $w_{\Lambda}$ and in how fast it changes. Since $\phi = \phi_0 e^{st}$, hence negative $s$ is preferred otherwise the field  evolves to super-Planckian regime.
 If considering that the present universe is expanding approximately like de-Sitter case, $r$ should be very small ($\sim 10^{-60}$) in equation (\ref{arrrr}). Therefore if considering scalar mass $m \sim M_{\rm P} = 1$, we would need $\k$ to be as large as $10^{60}$.


\section{Variation of gravitational constant}
A constraint to the model can be given by the measurement of gravitational constant variation. For example, the constraint with gravitational-wave standard sirens and supernovae is $\dot{G}/G|_{t_0} \lesssim 3 \times 10^{-12}\;\;\text{year}^{-1}$ at present time \cite{Zhao:2018gwk}. Constraint of the same order ($\dot{G}/G|_{t_0} \lesssim  10^{-12}\;\;\text{year}^{-1}$) is given by observations of pulsars \cite{Kaspi1994, Zhu:2018etc}, lunar laser ranging \cite{Williams2004} and Big Bang nucleosynthesis \cite{Copi2004, Cyburt2004}.
 The variation in our model is
\be
\f{\dot{G}_{\rm eff}}{G_{\rm eff}}   =  \f{- 24 \pi G \k \phi_0^2 s^3 e^{2 s t}}{1 + 12\pi G \k \phi_0^2 s^2 e^{2 s t}}\,.  \label{abcd}
\ee
As this is not consequence of equation (\ref{qpqp2}), we found that if without the scalar potential ($m = 0$ hence $s=0$), $\dot{G}_{\rm eff}/G_{\rm eff} = 0$, i.e. the gravitational constant is always constant.

\subsection{Variation of $G$: purely NMDC kinetic term ($\varepsilon = 0$)}
If there is only purely NMDC kinetic term, $\varepsilon = 0 $ and $s = -3r$ (a consequence of equation (\ref{qpqp2})). Using these relations in equation (\ref{abcd}) and
considering $t = t_0 = 0$ at present time,
\be
\l[\f{\dot{G}_{\rm eff}}{G_{\rm eff}}\r]_{t_0}   =  \f{   648  \pi G \k \phi_0^2 r^3 }{1 + 108\pi G \k \phi_0^2 r^2 }\,.  \label{abcded}
\ee
The variation diverges at singularity, $ \k_{{\rm s}, t_0} = (-108 \pi G \phi_0^2 r^2)^{-1}
 < 0\, $ which is $\k_{{\rm s}, t_0} \sim \, -7.4 \times 10^{118}$  (with $r \sim 10^{-60}$, $8 \pi G \equiv 1$, $m = 1$ and $\phi_0 = 1$).  The constraint $\dot{G}/G|_{t_0} \lesssim  10^{-12}\;\;\text{year}^{-1}$ limits the present-time $\k$ value to $ -7.4 \times 10^{118}  \lesssim  \k  \lesssim 7.4 \times 10^{118}  $ in Planck unit. However $\k \neq 0 $ must hold. The allowed range includes the singularity value of $\k$. That is $\k_{{\rm s}, t_0}$ is very slightly greater than $-7.4 \times 10^{118}$.
 Considering the equation of state, $\k > 0$ is favored, hence the result is concluded as $0 <  \k  \lesssim 7.4 \times 10^{118}$.

\subsection{Variation of $G$: canonical scalar field case ($\varepsilon = 1$)}
In finding variation of $G$ of this case, we do not use any consequence of the equation (\ref{qpqp2}), hence conditions $\k \neq 0, s \neq 0$ and $r \neq 0$ are not hold.   For the case $s = i m/\sqrt{\varepsilon}$, we have
\bea
&& \f{\dot{G}_{\rm eff}}{G_{\rm eff}}  =   \no \\
&&  \f{- \l( \f{ \a \k m^3}{\varepsilon\sqrt{\varepsilon}}\r) \sin\l(\f{2m t}{\sqrt{\varepsilon}}\r)   +
  i  \l( \f{\beta \k^2 m^5}{ \sqrt{\varepsilon} } \r) \l[ \f{2 \cos \l(\f{2m t}{\sqrt{\varepsilon}}\r)}{\a \k m^2 \varepsilon} - 1  \r]}{1 - \l(\f{\a \k m^2}{\varepsilon}\r)
  \cos\l(\f{2 m t}{\sqrt{\varepsilon}}\r) + \f{\b \k^2 m^4}{2} }\,,      \no  \\ \label{uhv}
\eea
and for the case $s = -i m/\sqrt{\varepsilon}$,
\bea
&& \f{\dot{G}_{\rm eff}}{G_{\rm eff}}  =   \no \\
&&  \f{- \l( \f{ \a \k m^3}{\varepsilon\sqrt{\varepsilon}}\r) \sin\l(\f{2m t}{\sqrt{\varepsilon}}\r)   +
  i  \l( \f{\beta \k^2 m^5}{ \sqrt{\varepsilon} } \r) \l[- \f{2 \cos \l(\f{2m t}{\sqrt{\varepsilon}}\r)}{\a \k m^2 \varepsilon} + 1  \r]}{1 - \l(\f{\a \k m^2}{\varepsilon}\r)
  \cos\l(\f{2 m t}{\sqrt{\varepsilon}}\r) + \f{\b \k^2 m^4}{2} }\,,  \no \\ \label{uhv2}
\eea
where  $\alpha \equiv 24  \pi G \phi_0^2  $  and  $\beta \equiv  {288 \pi^2 G^2 \phi_0^4 }    $.
For $\varepsilon = 1$, we see that real parts of equations (\ref{uhv}) and (\ref{uhv2}) are the same, i.e. the cases $s = i m $ and $s = -i m $ give the same real part of the equation of state. Considering present time, $t_0 = 0$, therefore
\be
 {\rm Re}\l[\f{\dot{G}_{\rm eff}}{G_{\rm eff}}\r]_{t_0} \; = \; 0\,.  \label{uhv3}
\ee
Hence in this case, at present time, there is no variation in the gravitational constant.

\subsection{Variation of $G$: phantom scalar field case ($\varepsilon = -1$)}
\begin{figure}[h!]
\begin{center}
\includegraphics[width=9.5cm,height=5.9cm,angle=0]{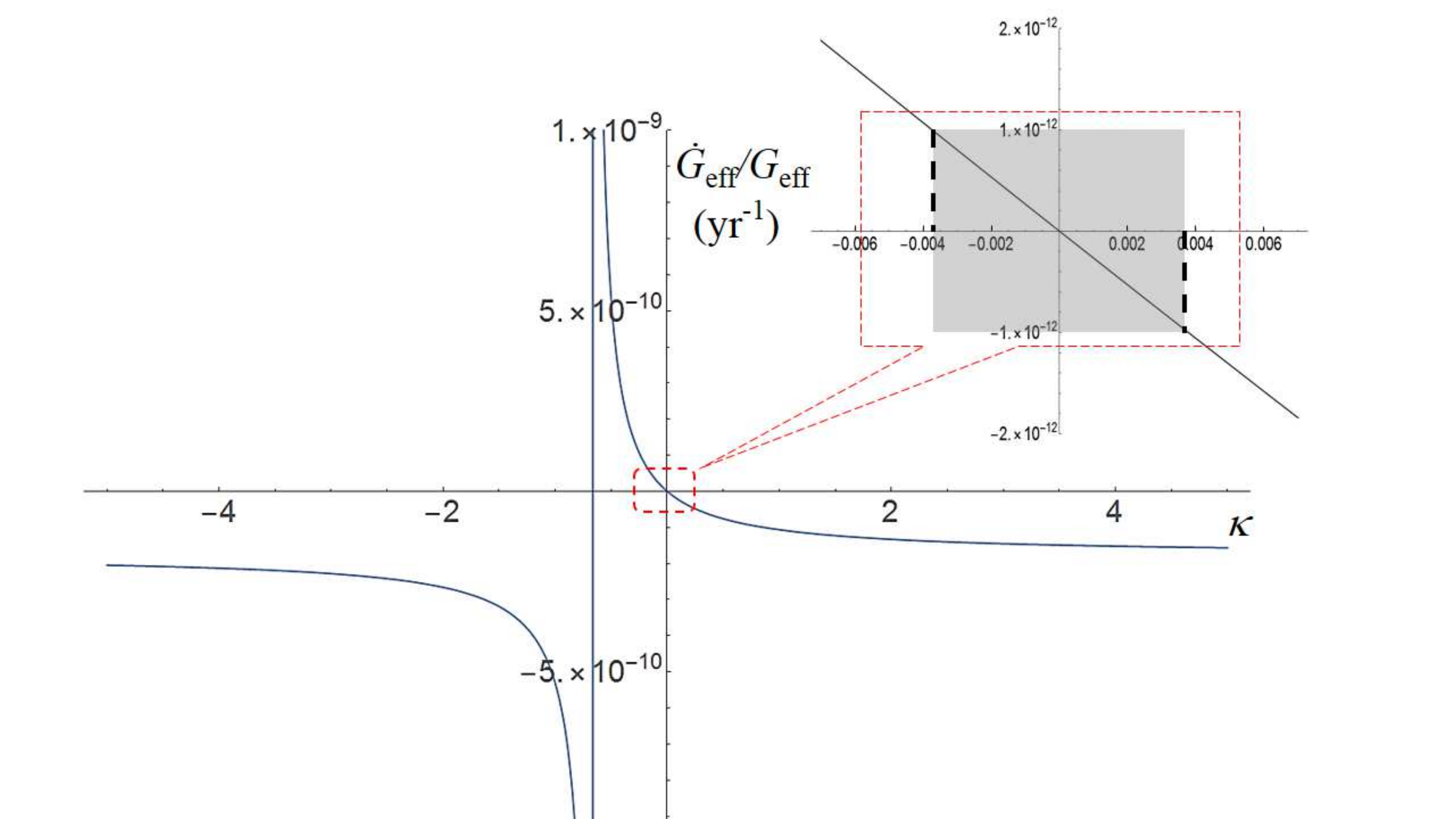}
\caption{Variation of $G_{\rm eff}$ in unit of $\text{year}^{-1}$ versus $\k$ in Planck unit for $\varepsilon = -1$ and $s= m$.  \label{Gdotem2}}
\includegraphics[width=9.5cm,height=5.9cm,angle=0]{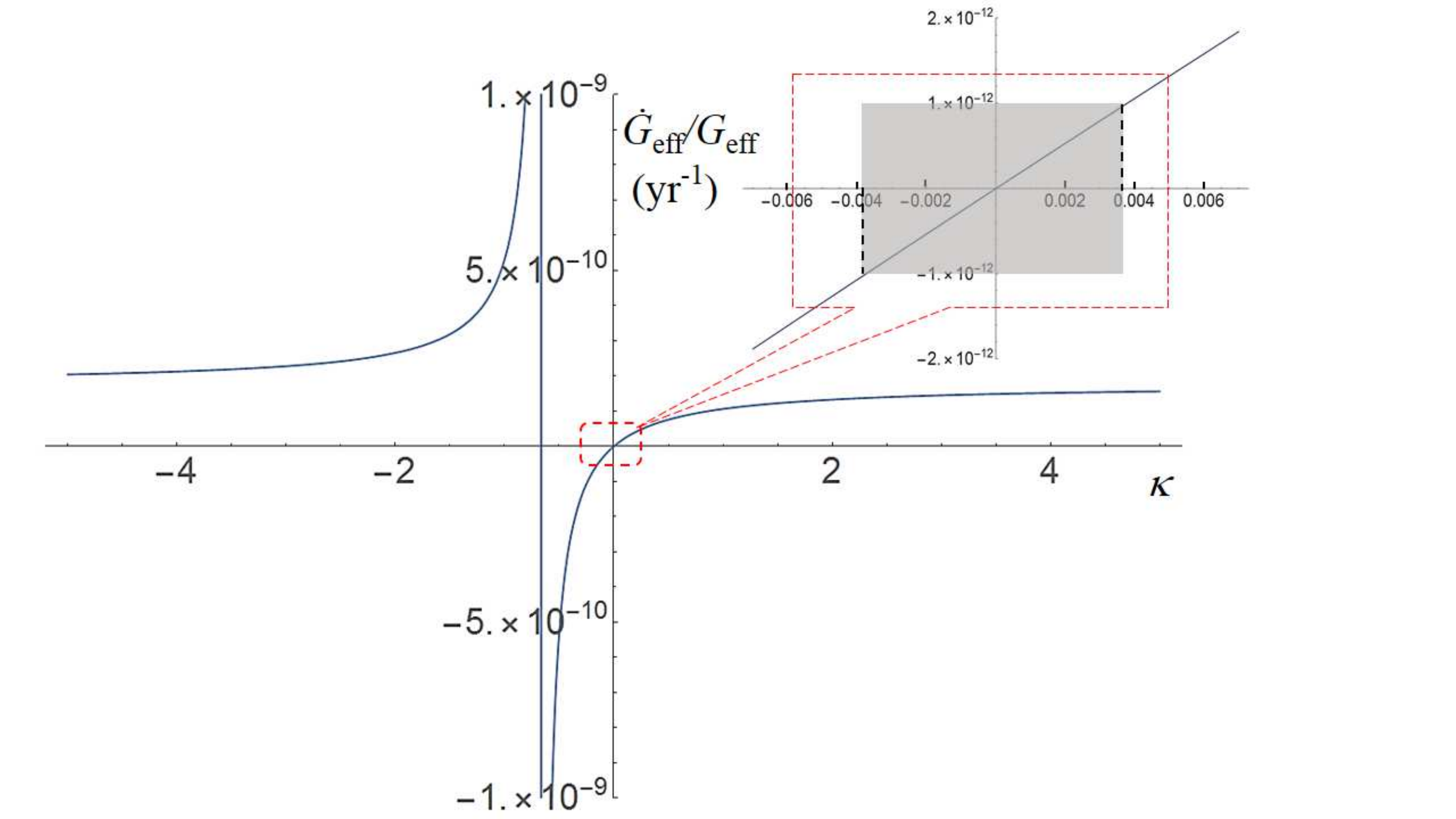}
\caption{Variation of $G_{\rm eff}$ in unit of $\text{year}^{-1}$ versus $\k$ in Planck unit for $\varepsilon = -1$ and $s = -m$. \label{Gdotem3}}\end{center}
  \end{figure}

In this section, consequence of the equation (\ref{qpqp2}) is neither used, therefore the conditions $\k \neq 0, s \neq 0$ and $r \neq 0$ are not hold.
Considering $\varepsilon = -1$ case, for $s = i m /\sqrt{\varepsilon} = m$, the variation is,
\bea
 \f{\dot{G}_{\rm eff}}{G_{\rm eff}}  =
  \f{-\a \k m^3 \sinh (2mt) - \b \k^2 m^5 \l[ \f{2 \cosh(2 mt)}{\a \k m^2} + 1 \r]      }{1 + \a \k m^2 \cosh(2mt) + \b \k^2 m^4/2 },
                            \label{uhv4}
\eea
and for $s = -i m /\sqrt{\varepsilon} = -m$,
\bea
 \f{\dot{G}_{\rm eff}}{G_{\rm eff}}  =
  \f{-\a \k m^3 \sinh (2mt) + \b \k^2 m^5 \l[ \f{2 \cosh(2 mt)}{\a \k m^2} + 1 \r]      }{1 + \a \k m^2 \cosh(2mt) + \b \k^2 m^4/2 }\,,
                            \label{uhv5}
\eea
where the distinct is the signs of the second term in each case.  At present time, we set $t_0 = 0$, hence
\bea
 \l[\f{\dot{G}_{\rm eff}}{G_{\rm eff}}\r]_{t_0}  =
  \f{\mp 288 \pi^2 G^2 \phi_0^4  \k^2 m^5 \l[ \f{2/3}{8 \pi G \phi_0^2 \k m^2} + 1 \r]      }{1 + 24 \pi G \phi_0^2  \k m^2   +   144 \pi^2 G^2 \phi_0^4 \k^2 m^4 }\,.
                            \label{uhv6}
\eea
The $\mp$ sign denotes the case $s = m$ and $s = -m$  respectively.
There is singularity at
\be
\k_{{\rm s}, t_0} = \f{-2/3}{8 \pi G \phi_0^2 m^2}   \,,
\ee
or, in Planck unit, it is $\k_{{\rm s}, t_0}  = -2/3$. We use the constraint $\dot{G}/G|_{t_0} \lesssim  10^{-12}\;\;\text{year}^{-1}$ to limit the value of $\k$. In the both cases  ($s = m$ and $s = -m$), we found that the constraints are the same, that is  $-0.0038 \lesssim \k \lesssim  0.0038$.
 We see how the ratio ${\dot{G}_{\rm eff}}/{G_{\rm eff}}$ changes with $\k$ for the phantom case in  Fig.
\ref{Gdotem2} and Fig. \ref{Gdotem3}. The grey shade denotes the constraint on $\k$ at present time.  This contradicts to the $w_{\Lambda}$ results in section \ref{sec_phantom} which requires the coupling to be super-Planckian or very large.

\section{Conclusions}
In this work, we study non-minimal derivative coupling (NMDC) to gravity in spatially flat FLRW universe in context of holographic dark energy. The theory has one free kinetic term and a kinetic NMDC term coupling to the Einstein tensor with constant coupling strength $\k$. The scalar potential is $V = (1/2)m^2 \phi^2$ and possibility of canonical field and phantom field is allowed. In the NMDC gravity, gravitational constant is modified with the NMDC kinetic term.
The limits of dark energy is introduced with the holographic IR cutoff. We takes cosmological length scale, that is, the Hubble horizon as our cutoff length scale. Hence dark energy density is, $\rho_{\Lambda} = 3 c^2H^2/(8 \pi G_{\rm eff})$. The dark energy density has then a combination of NMDC and holographic modification effects. Assuming exact solutions of the theory, we evaluate dark energy equation of state and the variation of gravitational constant of the theory in many possible cases of the solution. We put some constraints such that we can rule out
some cases of consideration. Conclusion for each possibility is
\begin{itemize}
\item {\bf Purely NMDC term:} The purely NMDC case result in  $w_{\Lambda} \longrightarrow -1$ at late time for $\k > 0$ while the $\k < 0$ case is not favored since $w_{\Lambda}$ either diverges from $[-1,1]$ or approaches $-1$ from the region with $w_{\Lambda} < -1$. Hence only $\k > 0$ case gives acceptable behavior.
    The scalar field evolves as $\phi = \phi_0 \exp(-3 r t)$ with $r$ could be as small as $\sim 10^{-60}$.
    In this case $w_{\Lambda}$ bends to $-1$ earlier for less $r$ as shown in Fig. \ref{fig2}.
      Gravitational constant variation constraint results that $-7.4 \times 10^{118} \lesssim \k  \lesssim 7.4 \times 10^{118}$  in Planck unit. The allowed range includes the singularity value $\k_{{\rm s}, t_0}$.
    Therefore the range $ 0 <  \k  \lesssim 7.4 \times 10^{118}$  is allowed for purely NMDC theory with potential $V=(1/2)m^2\phi^2$. However, to be realistic, it is possible that the coupling has sub-Planckian value, $0< \k < 1$.

\item {\bf Free kinetic and NMDC terms (canonical field):}
When allowing free kinetic term in the dynamics, as in equations (\ref{KGmod}) and ( \ref{Veffrr}), the free kinetic term takes part in the damping and the NMDC term takes part in modification of the {\it force}, i.e. the modifying slope of the potential with an extra piece ($   {6\kappa
H\dot{H}\dot{\phi}}/({\varepsilon - 3\kappa H^2})$).   The free kinetic term is considered in two possibilities, canonical or phantom fields. For the canonical field, some solutions of the theory (case 2) with $\k > 0 $ are favored. The favored case is the one with $w_{\Lambda} \rightarrow -1$ at late time and the one with $w_{\Lambda}$ that does not evolve from the phantom region in the past.
This
is with conditions,
$ s = -  i m$ with $r = [ 6  - m^2 \k (1  -    \sqrt{1  - {12}/({m^2 \kappa})})]^{1/2}/{\sqrt{18 \kappa}}$ and other conditions as stated in case 2.
As seen in Fig. \ref{fig8}, larger $\k$ makes $w_{\Lambda}$ approach $-1$ sooner.
The expansion function is $a = a_0 \exp(rt)$ with field oscillation, $\phi = \phi_0 \exp(-im t)$.
Singularity of $w_{\Lambda}$ at present is $\k_{\rm s, t_0} = 2/3$ and the singularity in $w_{\Lambda}$ at any time $t$ only happens in limited range $-2/3 \leq \k_{{\rm s}} \leq 2/3$.  Hence to avoid singularity at any time, $ \k > 2/3$ is suggested.  In this case, there is no variation in gravitational constant at present time.

\item {\bf Free kinetic and NMDC terms  (phantom field):}
For phantom field, $s = -m$ is preferred as it results that equation of state approaches $-1$ at late time. Negative $s$ is good because it prevents scalar field from increasing to super-Planckian regime. Positive coupling, $\k>0$ is required to avoid imaginary $r$ and imaginary mass $m$.
However the shortcomings are that the scalar mass $m$ and the coupling $\k$ are required to be super-Planckian while the constraint on gravitational constant variation puts a very narrow viable range of $\k$ as $-0.0038 \lesssim \k \lesssim  0.0038$.
\end{itemize}

In summary, $\k> 0$ is favored in all cases. The purely NMDC theory with the potential $V=(1/2)m^2\phi^2$ is favored with positive sub-Planckian coupling, satisfying $w_{\Lambda, t_0} \rightarrow -1$ and satisfying variation of gravitational constant constraint. The other viable case is the canonical field with NMDC term under the same potential with conditions stated in case 2 (section \ref{ohmygod}).  The phantom field case is disfavored.

As we mentioned in introduction (section \ref{sec_intro}), the canonical field NMDC theory with $V(\phi) = V_0 \phi^n$ can give acceleration phase for $n \leq 2$, with sub-Planckian $m$ and sub-Planckian positive $\k$ \cite{Skugoreva:2013ooa}. We see here that the acceleration phase is also possible with the addition of holographic dark energy idea (for $n = 2$).  It was claimed in \cite{Quiros:2017gsu} (see section \ref{sec_intro}) that the $\k > 0$ case could have classical (Laplacian) instability ($c_{\rm s}^2 < 0$). However  $c_{\rm s}^2$ depends on dynamical and kinematical variables, i.e. $\dot{\phi}, \ddot{\phi}, H$ and $\dot{H}$, hence inclusion of holographic cutoff should alter dynamics of the system and should change the range of $c_{\rm s}^2$ significantly. This is of interest for further study. We also notice that the NMDC coupling, $\k$ only affects $r$ (the exponent of the scale factor) but it does not contribute to any modification of $s$ which is the main character of scalar field evolution. Therefore further study can be investigated on some modification (with reasonable motivation) such that the NMDC coupling could contribute to both $r$ and $s$. Inclusion of barotropic density as dark matter can definitely allow NMDC effect to contribute to the expressions of $s$ and $r$. Therefore it is of further interests to perform dynamical analysis of such model with dark matter fluid inclusion.

\section*{Acknowledgments}
BG is supported by a TRF Basic Research Grant no. BRG6080003 (TRF Advanced Research Scholar) of the Thailand Research Fund and the Royal Society-Newton Advanced Fellowship (NAF-R2-180874).
BG and CK thank David Wands and Institute  of Cosmology and Gravitation, University of Portsmouth for hospitality during their academic visits. CM is supported by a Royal Golden Jubilee-ASEAN Ph.D. scholarship of the Thailand Research Fund. We thank Pichet Vanichchapongjaroen for useful comment.


\end{document}